\documentclass[12pt]{article}

\usepackage[english]{babel} 
\usepackage{natbib}
\usepackage{amssymb}
\usepackage{amsmath}
\usepackage{url}
\usepackage{txfonts}
\usepackage{rotating} 
\usepackage{xspace}
\usepackage{color}
\usepackage{lineno}
\usepackage{txfonts}
\usepackage{caption}
\usepackage{graphicx}
\usepackage{tabularx}
\usepackage[normalem]{ulem}

\begin{document}

\title{Micro-arcsecond structure of Sagittarius A* revealed by high-sensitivity 86 GHz VLBI observations}

   \author{\parbox{\textwidth}{
     Christiaan D. Brinkerink$^{1}$\thanks{E-mail:\texttt{c.brinkerink@astro.ru.nl}},
     Cornelia M\"uller $^{1,2}$\thanks{E-mail:\texttt{c.mueller@astro.ru.nl}},
     Heino D. Falcke $^{1,2}$,
     Sara Issaoun $^{1}$,
     Kazunori Akiyama $^{10,3,4}$,
     Geoffrey C. Bower $^{5}$,
     Thomas P. Krichbaum $^{2}$,
     Adam T. Deller $^{6}$,
     Edgar Castillo $^{7,8}$,
     Sheperd S. Doeleman $^{10,11}$,
     Raquel Fraga-Encinas $^{1}$,
     Ciriaco Goddi $^{1}$,
     Antonio Hern\'andez-G\'omez $^{12,13}$,
     David H. Hughes $^{8}$,
     Michael Kramer $^{2}$,
     Jonathan L\'eon-Tavares $^{8,14}$,
     Laurent Loinard $^{12,9}$,
     Alfredo Monta\~na $^{7,8}$,
     Monika Mo\'scibrodzka $^{1}$,
     Gisela N. Ortiz-Le\'on $^{2}$,
     David Sanchez-Arguelles $^{7}$,
     Remo P. J. Tilanus $^{1,15}$,
     Grant W. Wilson $^{16}$,
     J. Anton Zensus $^{2}$}}

\maketitle
   
\parbox{\textwidth}{\vspace{0.4cm}
$^{1}$ Department of Astrophysics/IMAPP, Radboud University, PO Box 9010, 6500 GL Nijmegen, The Netherlands
$^{2}$ Max-Planck-Institut f\"ur Radioastronomie, Auf dem H\"ugel 69, D-53121 Bonn, Germany
$^{3}$ National Radio Astronomy Observatory, 520 Edgemont Rd, Charlottesville, VA 22903, USA
$^{4}$ National Astronomical Observatory of Japan, 2-21-1 Osawa, Mitaka, Tokyo 181-8588, Japan
$^{5}$ Academia Sinica Institute of Astronomy and Astrophysics, 645 N. A'ohoku Pl., Hilo, HI 96720, USA
$^{6}$ Centre for Astrophysics and Supercomputing, Swinburne University of Technology, Mail Number H11, PO Box 218, Hawthorn, VIC 3122, Australia
$^{7}$ Consejo Nacional de Ciencia y Tecnolog\'ia, Av. Insurgentes Sur 1582, Col. Cr\'edito Constructor, Del. Benito Ju\'arez, C.P.: 03940, D.F., M\'exico
$^{8}$ Instituto Nacional de Astrof\'isica \'Optica y Electr\'onica (INAOE), Apartado Postal 51 y 216, 72000, Puebla, M\'exico
$^{9}$ Instituto de Astronomi\'ia, Universidad Nacional Auto\'onoma de Me\'exico, Apartado Postal 70-264, CdMx C.P. 04510, M\'exico
$^{10}$ Massachusetts Institute of Technology, Haystack Observatory, 99 Millstone Rd., Westford, MA 01886, USA
$^{11}$ Harvard Smithsonian Center for Astrophysics, 60 Garden Street, Cambridge, MA 02138, USA
$^{12}$ Instituto de Radioastronom\'ia y Astrof\'isica, Universidad Nacional Aut\'onoma de M\'exico, Morelia 58089, M\'exico
$^{13}$ IRAP, Universit\'e de Toulouse, CNRS, UPS, CNES, Toulouse, France
$^{14}$ Sterrenkundig Observatorium, Universiteit Gent, Krijgslaan 281-S9, B-9000 Gent, Belgium
$^{15}$ Leiden Observatory, Leiden University, P.O. Box 9513, 2300 RA Leiden, The Netherlands
$^{16}$ University of Massachusetts, Department of Astronomy, LGRT-B 619E, 710 North Pleasant Street, Amherst, MA01003-9305, USA
}

\date{\today}
 
\begin{abstract}
The compact radio source Sagittarius~A$^*$ (Sgr\,A$^*$\xspace) in the Galactic Center is the primary supermassive black hole candidate. General relativistic magnetohydrodynamical (GRMHD) simulations of the accretion flow around Sgr\,A$^*$\xspace predict the presence of sub-structure at observing wavelengths of $\sim 3$\,mm and below (frequencies of 86\,GHz and above). For very long baseline interferometry (VLBI) observations of Sgr\,A$^*$\xspace at this frequency the blurring effect of interstellar scattering becomes subdominant, and arrays such as the High Sensitivity Array (HSA) and the global mm-VLBI Array (GMVA) are now capable of resolving potential sub-structure in the source. Such investigations help to improve our understanding of the emission geometry of the mm-wave emission of Sgr\,A$^*$\xspace, which is crucial for constraining theoretical models and for providing a background to interpret 1\,mm VLBI data from the Event Horizon Telescope (EHT). We performed high-sensitivity very long baseline interferometry (VLBI) observations of Sgr\,A$^*$\xspace at 3\,mm using the Very Long Baseline Array (VLBA) and the Large Millimeter Telescope (LMT) in Mexico on two consecutive days in May 2015, with the second epoch including the Greenbank Telescope (GBT). We find an overall source geometry that matches previous findings very closely, showing a deviation in fitted model parameters less than 3\% over a time scale of weeks and suggesting a highly stable global source geometry over time. The reported sub-structure in the 3\,mm emission of Sgr\,A$^*$\xspace is consistent with theoretical expectations of refractive noise on long baselines. However, comparing our findings with recent results from 1\,mm and 7\,mm VLBI observations, which also show evidence for east-west asymmetry, an intrinsic origin cannot be excluded. Confirmation of persistent intrinsic substructure will require further VLBI observations spread out over multiple epochs.
\end{abstract}

\section{Introduction}

The radio source Sagittarius~A$^*$ (hereafter called Sgr\,A$^*$\xspace) is associated with the supermassive black hole (SMBH) located at the center of the Milky Way. It is the closest and best-constrained supermassive black hole candidate \citep{Ghez2008, Gillessen2009, Reid2009} with a mass of $M\sim4.1\times10^6 M_\odot$ at a distance of $\sim 8.1$\,kpc as recently determined to high accuracy by the GRAVITY experiment \citep{Gravity2018}. This translates into a Schwarzschild radius with an angular size of $\theta_\mathrm{R_S}\sim10\,\mu$as on the sky, while the angular size of its ``shadow'' -- the gravitationally lensed image of the event horizon -- is predicted to be $\sim50\,\mu$as \citep{Falcke2000}. Due to its proximity, Sgr\,A$^*$\xspace appears as the black hole with the largest angular size on the sky and is therefore the ideal laboratory for studying accretion physics and testing general relativity in the strong field regime \citep[see, e.g.,][for a review]{bhc_rev,FalckeMarkoff2013}.\\

\noindent
Radio observations of Sgr\,A$^*$\xspace have revealed a compact radio source with an optically thick spectrum up to mm-wavelengths. In the sub-mm band the spectrum shows a turnover and becomes optically thin. This sub-mm emission is coming from a compact region that is only a few Schwarzschild radii in size \citep[e.g.,][]{Falcke1998, Doeleman2008}. Very Long Baseline Interferometry (VLBI) observations can now achieve the required angular resolution down to a few tens of $\mu$as to resolve these innermost accretion structure close to the event horizon. The advantages in going to (sub-)mm wavelengths are 1.) to witness the transition from optically thin to thick emission, 2.) to improve the angular resolution and 3.) to minimize the effect of interstellar scattering. At longer radio wavelengths, interstellar scattering along our line of sight towards Sgr\,A$^*$\xspace prevents direct imaging of the intrinsic source structure and causes a ``blurring'' of the image that scales with wavelength squared \citep[e.g.,][]{Davies1976,Backer1978,Bower2014_pulsar}.\\

\noindent
The scatter-broadened image of Sgr\,A$^*$\xspace can be modeled by an elliptical Gaussian over a range of wavelengths. The measured scattered source geometry scales with $\lambda^2$ above observing wavelengths of $\sim$7\,mm \citep{Bower2006} following the relation: $({\theta_{\textrm{maj}} \over {1 \textrm{mas}}}) \times ({\theta_{\textrm{min}} \over {1 \textrm{mas}}}) = (1.31 \times 0.64) ({\lambda \over {\textrm{cm}}})^2$, with the major axis at a position angle $78^\circ$ east of north.  At shorter wavelengths this effect becomes subdominant, although refractive scattering could introduce stochastic fluctuations in the observed geometry that vary over time. This refractive noise can cause compact sub-structure in the emission, detectable with current VLBI arrays at higher frequencies \citep{JohnsonGwinn2015,Gwinn2014_scatteringsubstructure}.\\

\noindent
Due to major developments in receiver hardware and computing that have taken place over the past years, mm-VLBI experiments have gotten closer to revealing the intrinsic structure of Sgr\,A$^*$\xspace.  At 1.3\,mm (230\,GHz), the Event Horizon Telescope has resolved source structure close to the event horizon on scales of a few Schwarzschild radii \citep{Doeleman2008,Johnson2015}. Closure phase measurements over four years of observations have revealed a persistent East-West asymmetry in the 1.3\,mm emission of Sgr\,A$^*$\xspace \citep{Fish2016}. This observed structure and geometry seems intrinsic to the source and is already imposing strong constraints on GRMHD model parameters of Sgr\,A$^*$\xspace \citep{Broderick2016,Fraga-Encinas2016}. A more recently published analysis by \citep{Lu2018} of observations done at 230\,GHz including the APEX antenna reports the discovery of source substructure on even smaller scales of 20 to 30 $\mu$as that is unlikely to be caused by interstellar scattering effects.\\

\noindent
At 3.5\,mm (86\,GHz), the combined operation of the Large Millimeter Telescope (LMT, Mexico) and the Green Bank Telescope (GBT, USA) together with the Very Long Baseline Array (VLBA) significantly improves the (u,v)-coverage and array sensitivity beyond what is possible with the VLBA by itself. Closure phase analysis indicates an observational asymmetry in the 3\,mm emission \citep{Ortiz2016,Brinkerink2016}, which is consistent with apparent substructure introduced by interstellar scattering, although an interpretation in terms of intrinsic source structure cannot be excluded given the data obtained so far. \citet{Ortiz2016} reported on VLBA+LMT observations at 3.5\,mm detecting scattering sub-structure in the emission, similar to what was found at 1.3\,cm by \citet{Gwinn2014_scatteringsubstructure}. In \citet{Brinkerink2016}, using VLBA+LMT+GBT observations, we report on a significant asymmetry in the 3.5\,mm emission of Sgr\,A$^*$\xspace. Analyzing the VLBI closure phases, we find that a simple model with two point sources of unequal flux provides a good fit to the data. The secondary component is found to be located toward the East of the primary, however, the flux ratio of the two components is poorly constrained by the closure phase information. \\

\noindent
It remains unclear, however, whether this observed emission sub-structure at 3.5\,mm is intrinsic or arises from scattering. The body of VLBI observations reported so far cannot conclusively disentangle the two components. Time-resolved and multifrequency analysis of VLBI data can help. Besides the findings by \citet{Fish2016} at 1.3\,mm, \citet{Rauch2016} found a secondary off-core feature in the 7\,mm emission appearing shortly before a radio flare, which can be interpreted as an adiabatically expanding jet feature \citep[see also][]{Bower2004}.\\

\noindent
From elliptical fits to the observed geometry of the emission, the two-dimensional size of Sgr\,A$^*$\xspace at mm-wavelength can be derived as reported by \citet{Shen2005,Lu2011_sgra,Ortiz2016} at 3.5\,mm and \citet{Bower2004,Shen2006} at 7\,mm.  Using the known scattering kernel \citep{Bower2006,Bower2014}, this intrinsic size can be calculated from the measured size.  The most stringent constraint on the overall intrinsic source diameter has been determined using a circular Gaussian model for the observed 1.3\,mm emission \citep{Doeleman2008,Fish2011}, as at this observing frequency the scattering effect is less dominant.  More recent VLBI observations of Sgr\,A$^*$\xspace at 86\,GHz constrain the intrinsic, two-dimensional size of Sgr\,A$^*$\xspace to $(147\pm4)\mu\mathrm{as} \times (120\pm12)\mu\mathrm{as}$ \citep{Ortiz2016} under the assumption of a scattering model derived from \citet{Bower2006} and \citet{Psaltis2015}.\\

\noindent
High-resolution measurements of time-variable source structure in the infrared regime observed during Sgr\,A$^*$\xspace infrared flares have recently been published \citep{Gravity2018-2}, where spatial changes of the source geometry of Sgr\,A$^*$\xspace on timescales of less than 30 minutes are seen. These results suggest periodical motion of a bright source component located within $\sim$100 $\mu$as of the expected position of the supermassive black hole, with a corrrespondingly varying polarization direction. The variability timescale of Sgr\,A$^*$\xspace is expected to be significantly shorter at infrared wavelengths than at 3.5 and 1.3\,mm, as it is thought to be dominated by fast local variations in electron temperature rather than changes in the bulk accretion rate.

\noindent
All of these observations indicate that we start to unveil the presence of both stationary and time-variable sub-structure in the accretion flow around Sgr\,A$^*$\xspace, as expected by theoretical simulations \citep[e.g.,][]{Moscibrodzka2014}. In order to further put constraints on model parameters, higher-resolution and more sensitive mm-VLBI observations are required.  The analysis of closure quantities helps to determine source properties without being affected by station-based errors. Closure phases indicate asymmetry in the emission when significantly deviating from zero \citep[see, e.g.,][for the case of Sgr\,A$^*$\xspace]{Fish2016,Brinkerink2016}.  Closure amplitudes put constraints on the source size \citep[see, e.g.,][]{Ortiz2016,Bower2006,Bower2004}.  Imaging techniques are based on the closure quantities. Although mm-VLBI has a number of limitations, at $\gtrsim$3\,mm the current VLBI array configurations allow reconstructing the emission of Sgr\,A$^*$\xspace using standard hybrid imaging techniques \citep[][]{Lu2011_sgra,Rauch2016}.\\

\noindent
In this paper we follow up on our first analysis published in \citet{Brinkerink2016} (hereafter referred to as Paper~I). Here, we focus on the closure amplitude and imaging analysis of Sgr\,A$^*$\xspace at $\lambda=\:$3.5\,mm obtained with the VLBA and LMT on May 22nd, 2015 and VLBA, LMT, and GBT on May 23rd, 2015. In Section~\ref{sec:obs} we describe the observations and data reduction. Section~\ref{sec:results} discusses the results from imaging and closure amplitude analysis. In section 4, we present the results from a simultaneous fitting of the intrinsic size/frequency relation and the scattering relation for Sgr\,A$^*$\xspace, using the combined data from this work with earlier published results across a range of wavelengths. We conclude with a summary in Section~\ref{sec:summary}.\\

\section{Observations and Data Reduction}\label{sec:obs}

We performed 86\,GHz VLBI observations of Sgr\,A$^*$\xspace. Here we present the analysis of two datasets: one epoch using the VLBA (all 86\,GHz capable stations\footnote{Brewster (BR), Fort Davis (FD), Kitt Peak (KP), Los Alamos (LA), Mauna Kea (MK), North Liberty (NL), Owens Valley (OV) and Pie Town (PT)}) together with the LMT (project code: BF114A) on May 22nd, 2015, and one epoch using VLBA, LMT and GBT on May 23rd, 2015 (project code: BF114B). Both observations were observed in left-circular polarization mode only, at a center frequency of 86.068\,GHz and a sampling rate of 2 Gbps (512 MHz on-sky bandwidth). For fringe finding we used the primary calibrators 3C\,279 and 3C\,454.3 at the start and end of the track respectively. In between, the scans alternated every 5\,min between Sgr\,A$^*$\xspace and the secondary fringe finder NRAO\,530\xspace ([HB89] 1730-130) with short regular gaps (every $\sim$30 minutes) for pointing and longer GBT-only gaps every $\sim$4 hours for focusing.\\

\noindent
For fringe finding and initial calibration of both datasets, we used standard methods in \textit{AIPS} \citep{Greisen2003_AIPS} as described in Paper~I.  We first performed a manual phase-cal to determine the instrumental delay differences between IFs on a 5\,min scan of 3C\,454.3. After applying this solution to all data, the second FRING run gave us solutions for delay and rate (4\,min solution interval, with 2\,min subinterval) with a combined solution for all IFs. Using shorter solution intervals than the length we used here resulted in more failed, and therefore flagged, FRING solutions.  All telescopes yielded good delay/rate solutions for NRAO\,530\xspace.  For Sgr\,A$^*$\xspace, however, we found no FRING solutions on baselines to MK (using a limiting value for the Signal-to-Noise Ratio (S/N) of 4.3), but all other baselines yielded clear detections.\\

\noindent
Amplitude calibration in \textit{AIPS} was performed using a-priori information on weather conditions and gain-elevation curves for each station. In the cases of the LMT and the GBT, system temperature measurements and gain curves were imported separately as they were not included in the a-priori calibration information provided by the correlator pipeline. We solved for (and applied) atmospheric opacity corrections using the \textit{AIPS} task \texttt{APCAL}. To prepare for the remaining amplitude corrections, the data were then IF-averaged into a single IF and exported to \textit{Difmap} \citep{Shepherd1997}.\\

\begin{figure*}
\includegraphics[width=0.5\textwidth]{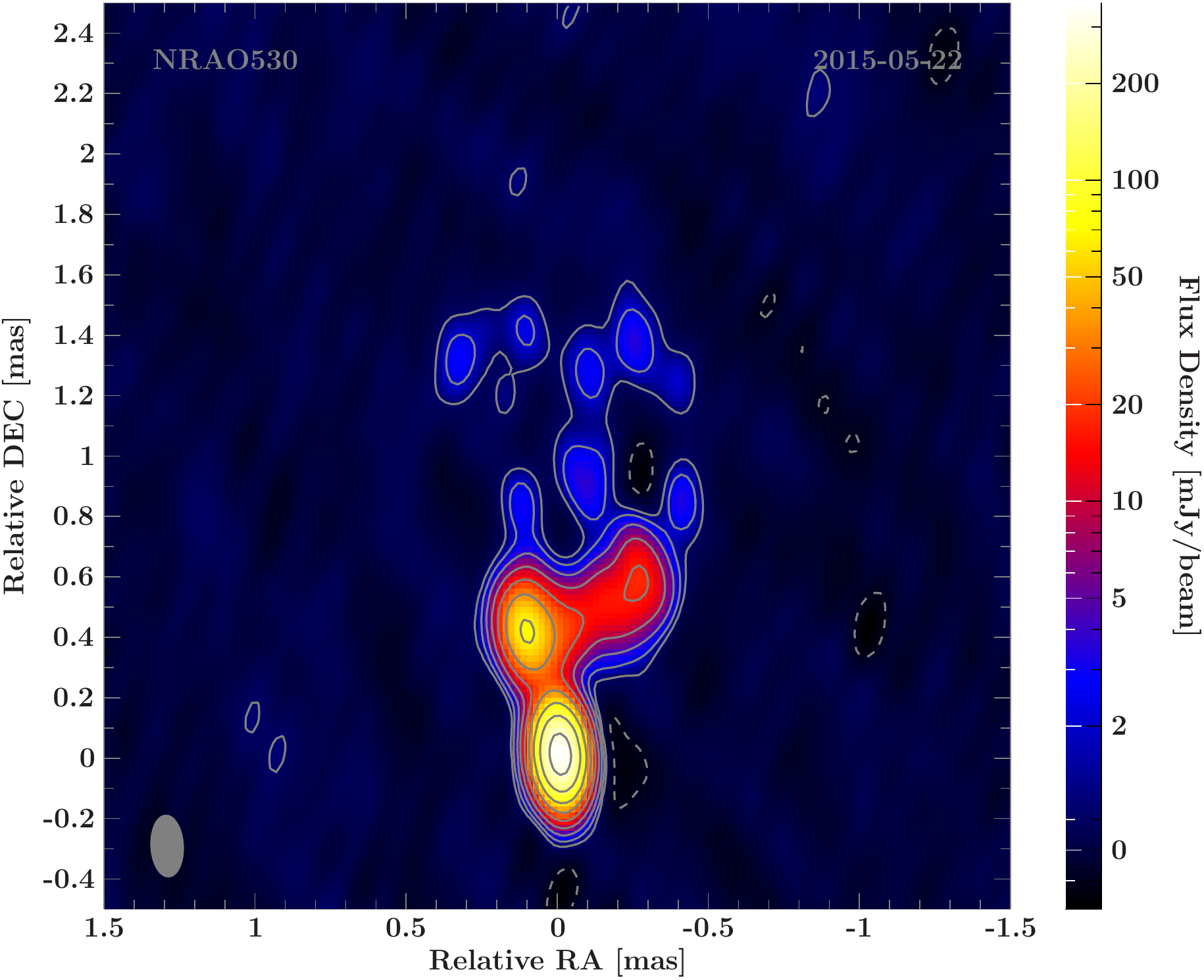}
\includegraphics[width=0.5\textwidth]{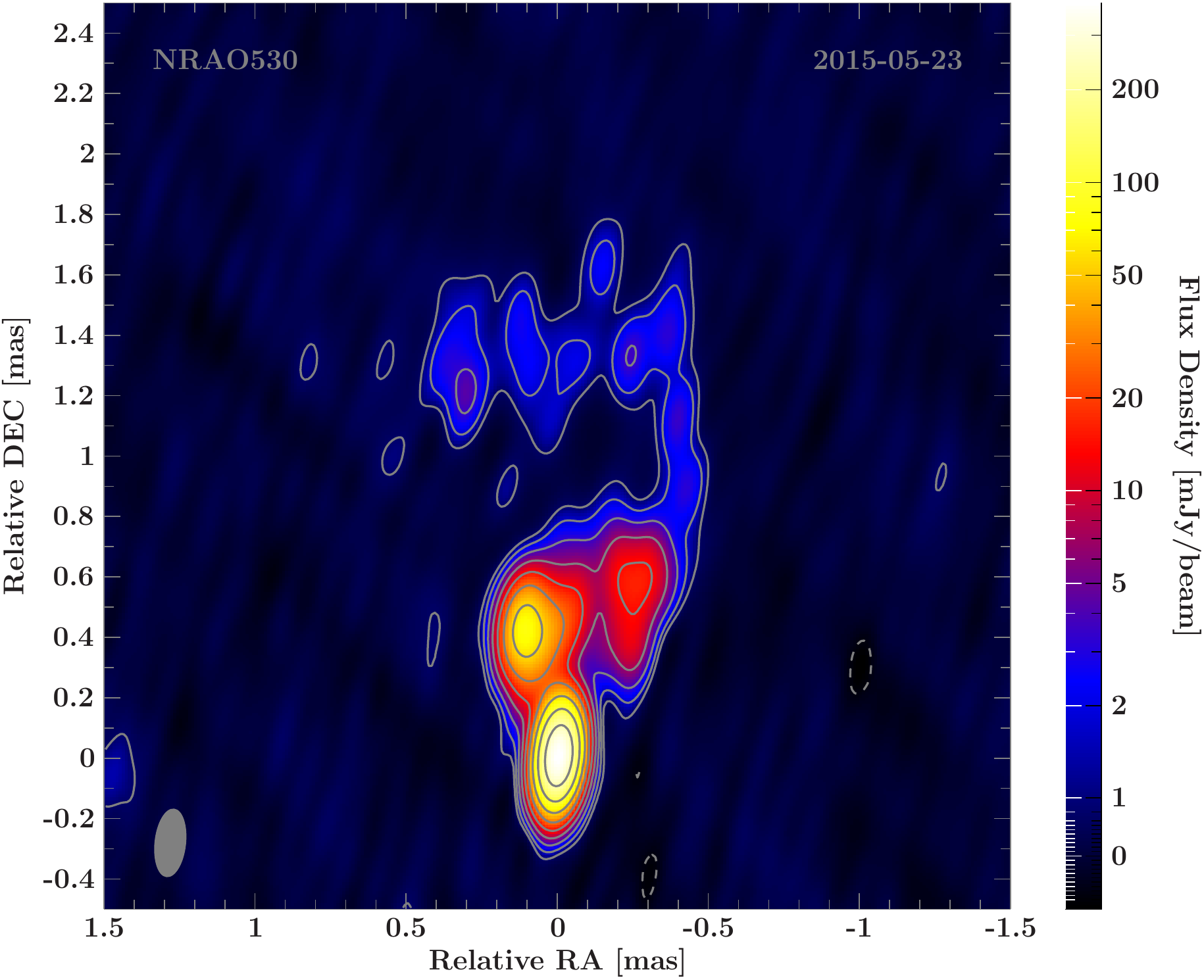}
 \caption{Naturally weighted 86\,GHz images of NRAO\,530\xspace. \textit{Left:} using data of project BF114A (2015-05-22) with VLBA and LMT.  \textit{Right:} using data of project BF114B (2015-05-23) with VLBA, LMT and GBT. The contours indicate the flux density level (dashed-gray contours are negative), scaled logarithmically and separated by a factor of 2, with the lowest level set to the 3$\sigma$-noise level. The synthesized array beam is shown as a gray ellipse in the lower left corner. Image parameters are listed in Table~\ref{table:images}. }
\label{fig:nrao}
\end{figure*}

\noindent
The quality of millimeter-VLBI observations is in practice limited by a number of potential error contributions \citep[cf.,][]{MartiVidal2012}: atmospheric opacity and turbulence, and telescope issues (e.g., pointing errors). In the case of Sgr\,A$^*$\xspace, the low elevation of the source for Northern Hemisphere telescopes requires a careful calibration strategy as loss of phase coherence needs to be avoided, and atmospheric delay and opacity can fluctuate relatively quickly at 86\,GHz with a coherence timescale typically in the range of 10 to 20 seconds.\\ We therefore used NRAO\,530\xspace as a test source to get a handle on the uncertainties and potential errors in the data. NRAO\,530\xspace has been extensively studied with VLBI at different wavelengths \citep[e.g.,][]{Lu2011_sgra,Lu2011_nrao530,An2013} and is regularly monitored with the VLBA at 43\,GHz in the framework of the Boston University Blazar Monitoring Program\footnote{url{http://www.bu.edu/blazars/VLBAproject.html}}, providing a good body of background knowledge on the source structure and evolution. For this source, we performed standard hybrid mapping in \textit{Difmap}. Using an iterative self-calibration procedure with progressively decreasing solution intervals, we obtained stable CLEAN images with sidelobes successfully removed. Careful flagging was applied to remove low-S/N and bad data points. Figure~\ref{fig:nrao} shows the naturally weighted CLEAN images for both datasets. Table~\ref{table:images} includes the corresponding image parameters.  The overall source structure is comparable between the two tracks, and the total recovered flux density in both images differs by less than 10\%. With ALMA-only flux measurements of NRAO\,530\xspace, a significantly higher total flux of 2.21\,Jy at band 3 (91.5\,GHz, ALMA Calibrator database, May 25, 2015) was measured. The difference with the flux we measured from the VLBI observations is likely due to a significant contribution from large-scale structure which is resolved out on VLBI baselines. Because the GBT and the LMT have adaptive dish surfaces, their gain factors can be time-variable. As such their gain curves are not fixed over time, and so additional and more accurate amplitude calibration in \textit{Difmap} was required for baselines to these stations. The imaging procedure started with an initial source model based on VLBA-only data, which allowed us to obtain further amplitude correction factors for the LMT of 1.47 (BF114A) and 1.14 (BF114B), and for the GBT of 0.54 (both tracks). Gain correction factors for the VLBA stations were of the order of $\lesssim$20\%.\\

\noindent
Due to the gain uncertainty for the GBT and the LMT for the reason mentioned above, amplitude calibration for Sgr\,A$^*$\xspace required a further step beyond the initial propagation of gain solutions from scans on NRAO\,530\xspace to scans on Sgr\,A$^*$\xspace. This calibration step was performed by taking the Sgr\,A$^*$\xspace visibility amplitudes from the short baselines between the South-Western VLBA stations (KP, FD, PT, OV) and using an initial model fit of a single Gaussian component to these VLBA-only baselines. Due to the low maximum elevation of Sgr\,A$^*$\xspace (it appears at $\sim$16 degrees lower elevation than NRAO\,530\xspace at transit), the amplitude correction factors for the VLBA are typically larger for Sgr\,A$^*$\xspace than for NRAO\,530\xspace but still agree with the factors of the corresponding NRAO\,530\xspace observations within $\lesssim 30$\% (except for the most Northern stations BR and NL), comparable to findings by \citet{Lu2011_sgra}. Analogously to the data reduction steps taken for NRAO\,530\xspace, we used this initial source model to perform additional amplitude calibration for the GBT and the LMT. After this first round of amplitude self-calibration, iterative mapping and self-calibration was performed (see Sect.~\ref{sec:imaging}).\\

\section{Results}\label{sec:results}

Following the closure phase analysis in Paper~I, we now study the source geometry and size using hybrid imaging (Sect.~\ref{sec:imaging}) and closure amplitudes (Sect.~\ref{sec:closureampl}). In Paper~I, where we studied the closure phase distribution to look for source asymmetry, we concentrated only on the more sensitive dataset including VLBA+LMT+GBT (project code: BF114B), while in this paper we also include the VLBA+LMT dataset (project code: BF114A).

\subsection{Mapping and Self-calibration of Sgr\,A$^*$\xspace}\label{sec:imaging}

\begin{figure*}
\includegraphics[width=0.5\textwidth]{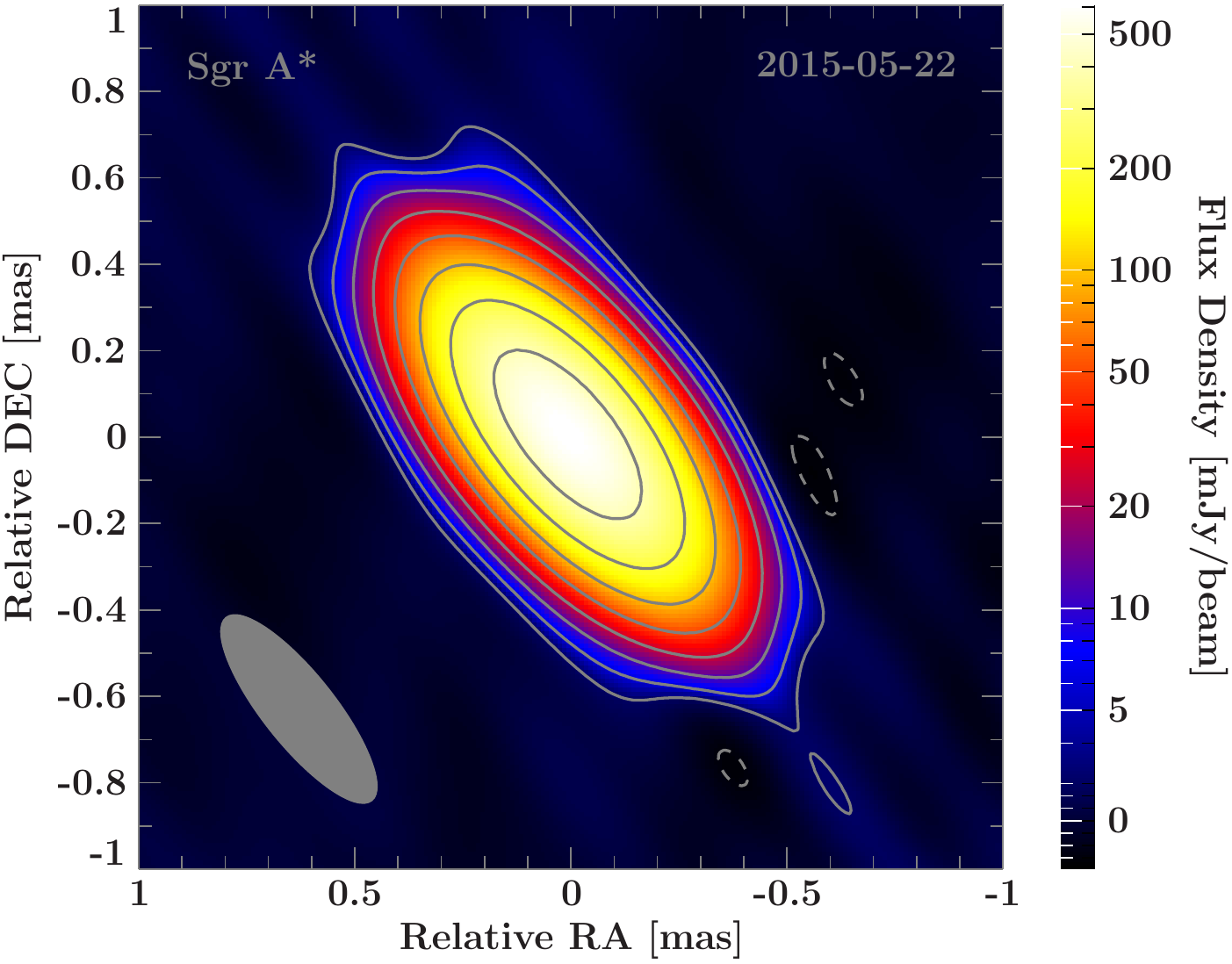}
\includegraphics[width=0.5\textwidth]{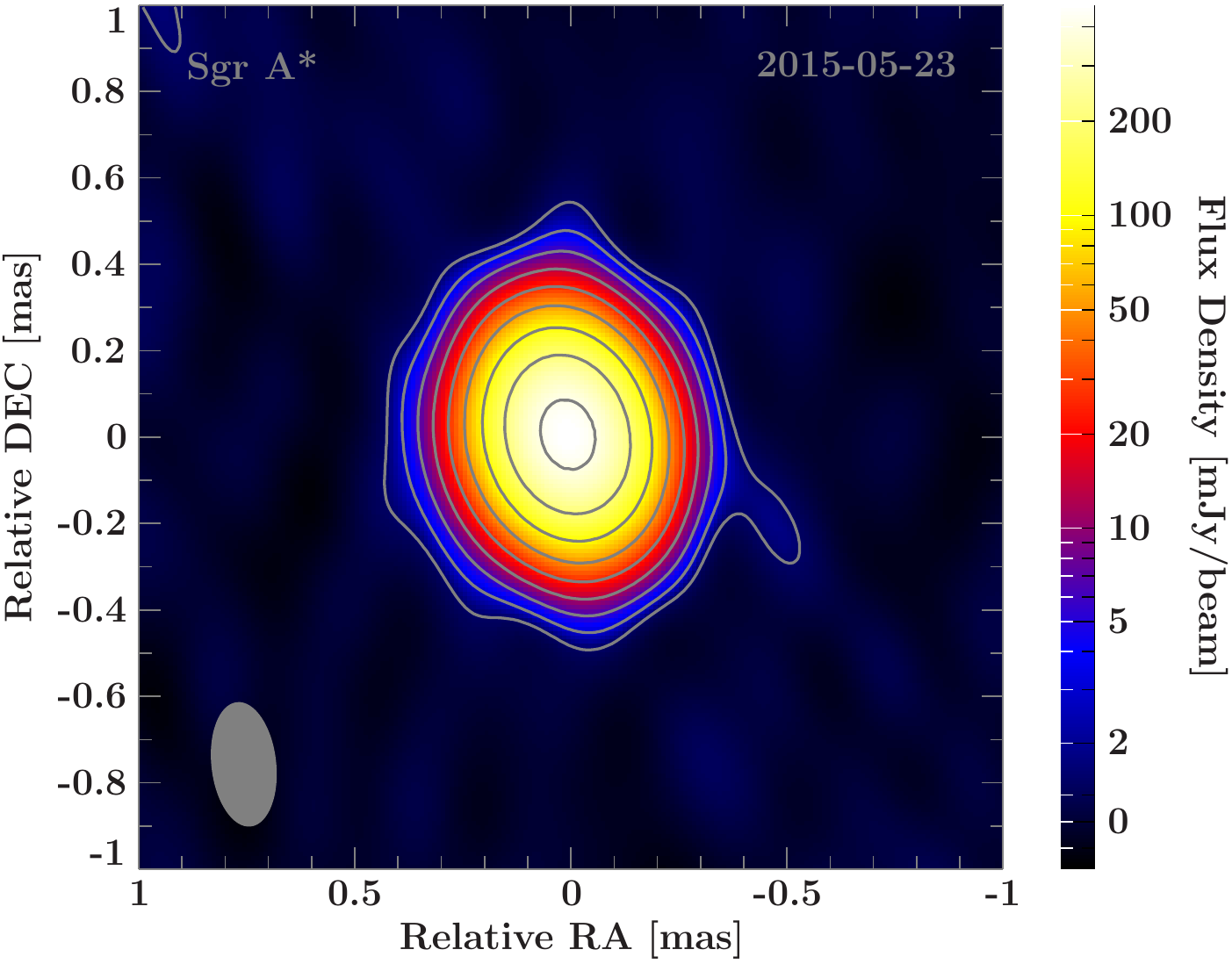}
\includegraphics[width=0.5\textwidth]{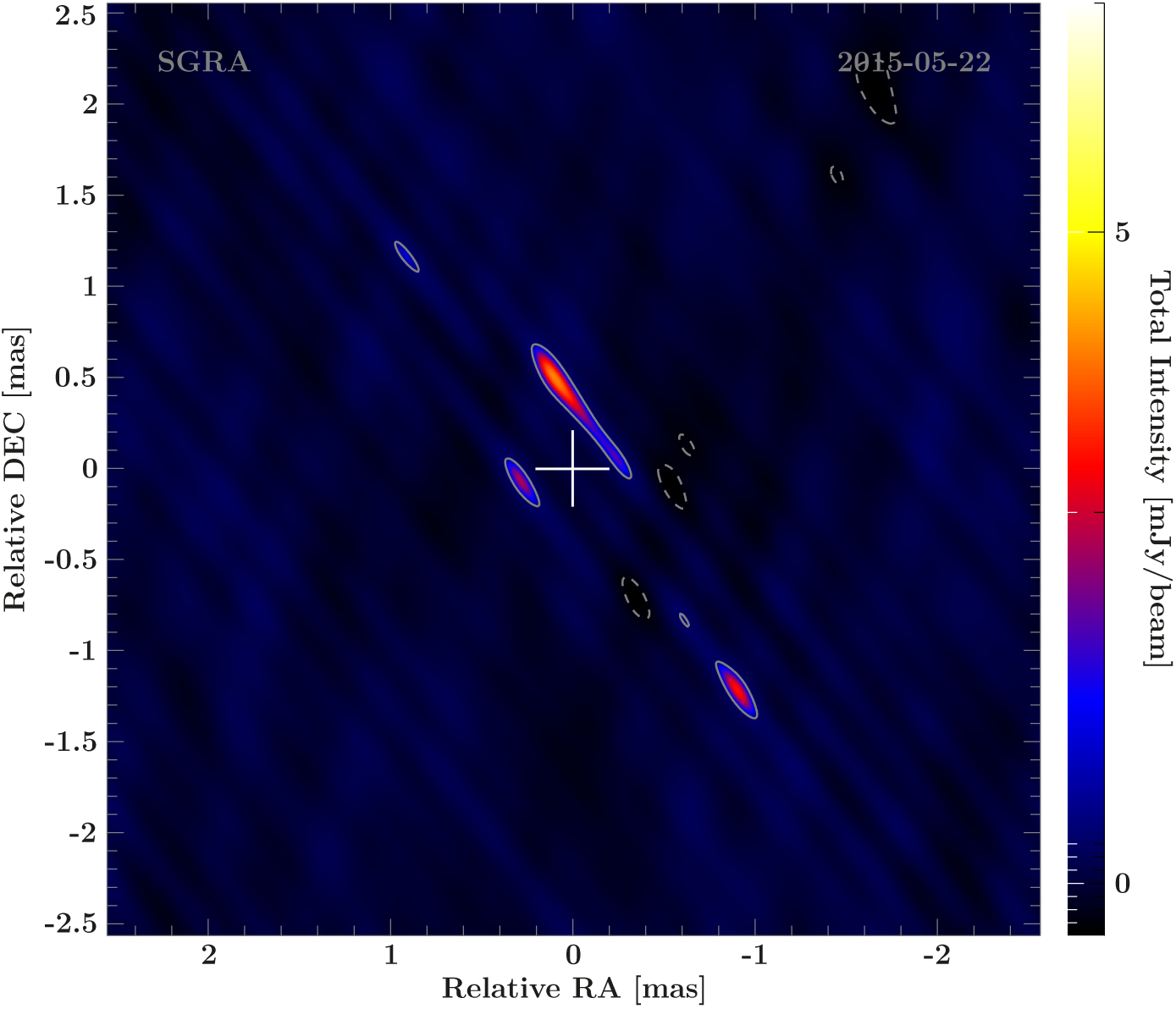}
\includegraphics[width=0.5\textwidth]{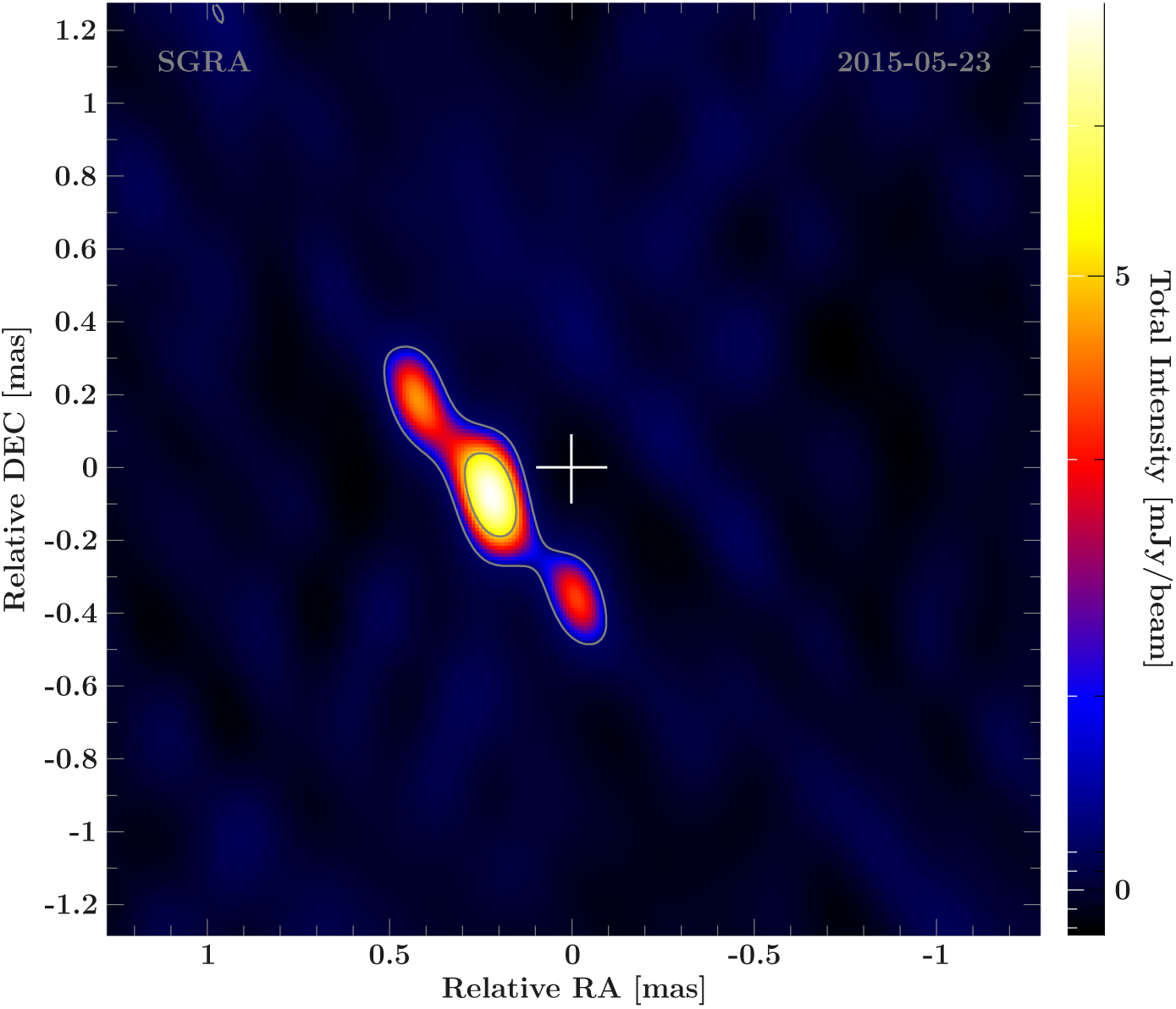}
\caption{Results of hybrid mapping of Sgr~A* at 3\,mm. \textit{Top left:} Beam-convolved image from the dataset of project BF114A (2015-05-22) using VLBA and LMT. \textit{Top right:} Beam-convolved image from  the dataset of project BF114B (2015-05-23) using VLBA, LMT and GBT. The contours indicate the flux density level (dashed-gray contours are negative), scaled logarithmically and separated by a factor of 2, with the lowest level set to the 3$\sigma$-noise level. \textit{Bottom left:} Residual map of Sgr\,A$^*$\xspace after primary component subtraction from the BF114A dataset, using natural weighting. No clear pattern is seen in the residual image. \textit{Bottom right:} Natural-weighted Residual map for Sgr A*, epoch B, after subtraction of the best-fitting 2D Gaussian source component. The remaining excess flux towards the East is highly concentrated and clearly present. Both residual images use a cross to indicate the center of the primary (subtracted) component on the sky. Image and model parameters are listed in Table~\ref{table:images} and~\ref{table:modelfits}, respectively.}
\label{fig:sgra}
\end{figure*}

After amplitude correction factors were applied (as explained in Sect.~\ref{sec:obs}), we performed an iterative mapping and self-calibration procedure including careful flagging of the Sgr\,A$^*$\xspace dataset. Amplitude and phase self-calibration were applied using increasingly shorter timesteps and natural weighting. We deconvolved the image for both datasets by using elliptical Gaussian model components, since the CLEAN algorithm has difficulty fitting the visibilities when it uses point sources. Table~\ref{table:modelfits} gives the best-fit parameteres from this approach. Figure~\ref{fig:sgra} shows both of the resulting images convolved with the clean beam.\\

\noindent
As shown by, e.g., \citet{Bower2014}, when self-calibrating, the derived model can depend on the initial self-calibration model chosen for a single iteration, if the $\chi^2$-landscape has complex structure. Furthermore, as also noted by \citet{Ortiz2016}, the resulting uncertainties on the model parameters are often underestimated, if they are based solely on the self-calibration solution. To assess the true errors, the uncertainties on the gain solutions must also be taken into account.\\

\begin{figure}
\includegraphics[width=\textwidth]{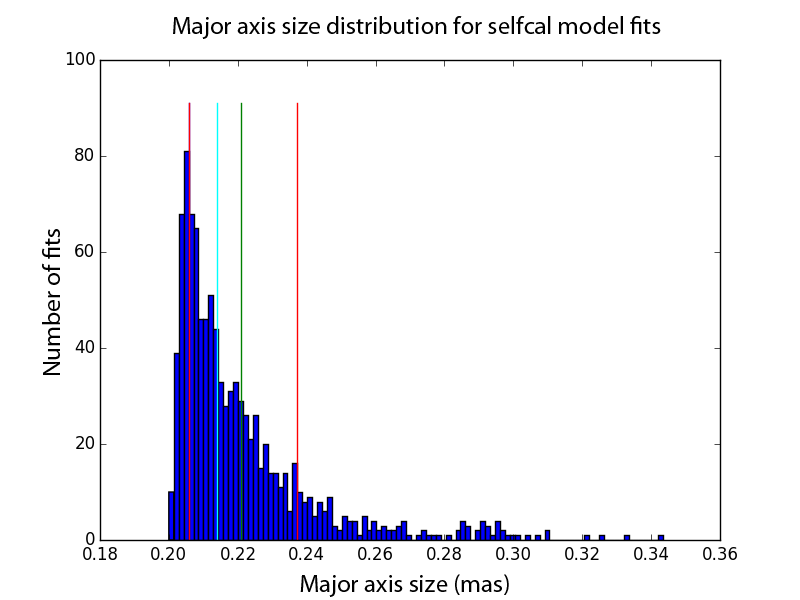}
\caption{The distribution of major axis sizes arising from 1000 selfcal runs in which each initial model parameter was varied according to a Gaussian distribution with a width of 30\% of the nominal parameter value. The resulting distribution of sizes shows a clear skew, with most results clustering close to a minimum cutoff value of $200\,\mu$as. The coloured lines indicate the mean (green), the median (cyan) and the statistical 1-$\sigma$ errors (red).}
\label{fig:selfcaldistribution}
\end{figure}

\noindent
Therefore, we tested the robustness of the final model, i.e., the dependence of the self-calibration steps on input models, described as follows. We evaluated conservative uncertainties on the model parameters of the elliptical Gaussian by using different starting parameters for the iterative self-calibration procedure, where all starting model parameters were individually varied by up to 30\%, to check the convergence on the same solution.  We generated 1000 random starting models to perform the initial amplitude self-calibration (Sect.~\ref{sec:obs}). The starting model always consists of an elliptical Gaussian. Each of its parameters (flux, major axis, axial ratio, position angle) was drawn from a normal distribution around the initial model. Using these input models, iterative self-calibration steps were applied and the resulting distribution of the model parameter was examined. For an illustration of the observed distribution of the major axis size, please see Figure \ref{fig:selfcaldistribution}.\\

\noindent
As expected, we find a strong correlation between input model flux density and final flux density of the Gaussian model components. Therefore, to constrain the flux of Sgr\,A$^*$\xspace, we primarily used the fluxes on short VLBA baselines as explained in Sect.~\ref{sec:obs}. For both NRAO\,530\xspace and Sgr\,A$^*$\xspace, we find less than 10\% total flux density difference between our two consecutive epochs.\\

\noindent
We find that the model converges onto values for major axis and axial ratio (or alternatively minor axis) that show a spread of about 10\%.  The position angle uncertainty is constrained to $\pm20^\circ$. Note that this analysis shows that the distribution for the major axis in BF114B is skewed, having an average of $222\,\mu$as, a median of $215\,\mu$as and a mode of $205\,\mu$as. The resulting major axis distribution also has a hard lower bound at $\sim\,200\mu$as. This skewed distribution of parameters from selfcal suggests that there are multiple local minima in the $\chi^2$-landscape that make the model parameters come out differently between iterations, and is therefore of limited value in determining source size uncertainties. We have therefore used closure amplitude analysis to verify this estimate of the source size and provide more accurate uncertainties, and this process is described in the next section.\\

\noindent
We find that for BF114A (VLBA+LMT), one single Gaussian component is sufficient to model the data (see Figure \ref{fig:sgra}, bottom left). For the BF114B dataset with higher sensitivity due to the inclusion of the GBT, the model fitting with one Gaussian component shows a significant excess of flux towards the South-West in the residual map (see Figure \ref{fig:sgra}, bottom right). Modeling this feature with a circular Gaussian component yields a flux density excess of $\sim10$\,mJy (i.e., approximately 1\% of the total flux) at $\Delta$RA$\sim0.23$ mas, $\Delta$DEC$\sim-0.05$ mas from the phase center. Including this second component in the modelfit, results in a smooth residual map (with RMS$\sim$0.5\,mJy). We checked the reliability of this feature using the same method as described above, where a range of initial model parameters was used as input for a selfcalibration step that resulted in a distribution of best-fit model parameters. We find that the position of the residual emission is well-constrained and independent from the self-calibration starting parameters. The BF114A dataset, however, does not show such clear and unambiguous residual emission.\\

We have tested the compatibility of the BF114A dataset with the source model we find for BF114B. Subtracting the full 2-component BF114B source model from the calibrated BF114A data and looking at the residual map, we see an enhanced overall noise level and no clear evidence of missing flux at the position of the secondary component. We further performed a separate amplitude and phase selfcalibration of the BF114A data using the BF114B source model, and inspected the residual map after subtraction of only the main source component of the BF114B model. In this residual map, we do see an enhancement of flux density at the position of the secondary component, but it is not as strong as the secondary component of the source model ($\sim5$\,mJy versus 10\,mJy for the model). We also see apparent flux density enhancements of similar strength at other positions close to the phase center. We therefore conclude that the BF114A $(u,v)$-coverage and sensitivity are not sufficient to provide a clear measurement of the secondary source component as seen for the BF114B epoch. Given that the detectability of the secondary component is so marginal for BF114A, we cannot determine whether the asymmetry we see in the BF114B epoch is a feature which persisted over the two epochs or a transient feature that was not present in the earlier epoch.\\

\noindent
We emphasize that the asymmetric feature we see in the Sgr\,A$^*$\xspace emission when imaging BF114B was already suggested by our analysis of the closure phases of the BF114B dataset (Paper~I). We found that a model consisting of two point sources results in a significantly better fit to the closure phases, with the weaker component being located East of the primary. However, the flux ratio of the two components was left poorly constrained, resulting in $\chi^2$ minima at flux ratios of 0.03, 0.11, and 0.70. In the current analysis, by using the full visibility data and fitting Gaussian components instead of point sources, we can constrain the flux ratio to $\sim$0.01. The low flux density of this secondary source component compared to the main source component makes it difficult to detect this source feature upon direct inspection of the visibility amplitudes as a function of baseline length. However, with model fitting it becomes clear that a single Gaussian component systematically underfits the amplitude trends of the data. We have thus seen evidence for this component independently in both the closure phases (Paper~1) and the visibility amplitudes (this work).\\

\noindent
It remains unclear whether this substructure in the 3\,mm emission of Sgr\,A$^*$\xspace is intrinsic or induced by refractive scattering. On long baselines, refractive scattering can introduce small-scale substructure in the ensemble-averaged image \citep{JohnsonGwinn2015}. This effect strongly depends on the intrinsic source size and geometry. A larger source size will show smaller geometrical aberration from scattering compared to a point source, as different parts of the source image are refracted in independent ways that tend to partially cancel out any changes in overall structure. At $\lambda\sim5\,$mm where the intrinsic source size of Sgr\,A$^*$\xspace becomes comparable to the angular broadening, this effect is most distinct \citep{JohnsonGwinn2015}. \citet{Gwinn2014_scatteringsubstructure} reported on the detection of scattering substructure in the 1.3\,cm emission of Sgr\,A$^*$\xspace. Assuming a Kolmogorov spectrum of the turbulence, the authors expect refractive scintillation to lead to the flux density measured on a 3000 km east-west baseline to vary with an RMS of 10-15 mJy. Similarly, \citet{Ortiz2016} show that refractive effects can cause substructure in 3\,mm images, with a RMS flux modulation of 6.6\,\% and an evolution timescale of about two weeks. Taking these considerations into account, this substructure detected at long baselines in our 3\,mm datasets would be consistent with scattering noise. However, given the more significant detection in the dataset involving the GBT and LMT, a contribution of intrinsic substructure cannot be excluded. We discuss more implications in Sect.~\ref{sec:summary}.\\

\begin{table*}
\centering
\scriptsize
\addtolength{\leftskip} {-2cm}
\addtolength{\rightskip}{-2cm}
\caption{Image \& observational parameters (natural weighting) \label{table:images}}
\begin{tabular}{cccccc}
\hline\hline
Date (Project ID) & Source & Array configuration$^a$ & Beam & Beam PA & RMS\\
yyyy-mm-dd &  & & [mas] & &[Jy/beam]\\
\hline
2015-05-22 (BF114A) & NRAO\,530\xspace & VLBA+LMT & 0.107$\times$0.204&$3.0^\circ$ &0.0004\\
2015-05-23 (BF114B) &NRAO\,530\xspace & VLBA+LMT+GBT & 0.100$\times$0.225&$-6.3^\circ$&0.0003\\
\hline
2015-05-22 (BF114A) & Sgr\,A$^*$\xspace & VLBA+LMT &0.541$\times$0.165 & $38.5^\circ$ & 0.0010\\
2015-05-23 (BF114B) &Sgr\,A$^*$\xspace & VLBA+LMT+GBT & 0.147$\times$0.286 & $6.4^\circ$ & 0.0005\\
\hline
\end{tabular}\\
$^a$ for VLBA: Brewster (BR), Fort Davis (FD), Kitt Peak (KP), Los Alamos (LA), Mauna Kea (MK), North Liberty (NL), OVRO (OV) and Pie Town (PT). Note that for Sgr\,A$^*$\xspace no fringes were detected to MK, which results in a larger beam size for Sgr\,A$^*$\xspace than for NRAO\,530\xspace.
\end{table*}

\begin{table*}
\centering
\scriptsize
\addtolength{\leftskip} {-2cm}
\addtolength{\rightskip}{-2cm}
\caption{Parameters of model components from self-calibration\textbf{$^a$}\label{table:modelfits}}
\begin{tabular}{lccccccc}
\hline\hline
Date (Project ID)& $S$ & $b_\mathrm{maj}$ & ratio &
   $b_\mathrm{min}$  & PA  \\
yyyy-mm-dd & [Jy] & [$\mu$as] & -- &[$\mu$as]  & [deg]\\
\hline
2015-05-22 (BF114A) & $1.02\pm0.1$ & $227.0$ & $0.85$ & $193.0$ & $56.4$\\
2015-05-23 (BF114B) & $0.95\pm0.1$ & $215.3$ & $0.77$ & $165.8$ &  $76.5$\\
\hline
\end{tabular}\\
$^a$Note: major/minor axis uncertainties are of the order of 10\%. The PA is constrained to within $15^\circ$ (BF114A) and $12^\circ$ (BF114B). See Sect.~\ref{sec:imaging} for more details.
\end{table*}

\subsection{Constraining the size of Sgr\,A$^*$\xspace using closure amplitudes}\label{sec:closureampl}

Closure quantities are robust interferometric observables which are not affected by any station-based error such as noise due to weather, atmosphere or receiver performance. As one example of a closure quantity, the closure phase is defined as the sum of visibility phases around a closed loop, i.e., at least a triangle of stations. We discussed the closure phase analysis of the Sgr\,A$^*$\xspace dataset BF144B in Paper~I in detail. Here, instead of closure phases, we focus on the closure amplitude analysis of both datasets. The closure amplitude is defined as $|V_{ij} V_{kl}|/|V_{ik} V_{jl} |$, for a quadrangle of stations $i,j,k,l$ and with $V_{ij}$ denoting the complex visibility on the baseline between stations $i$ and $j$. Using measurements of this quantity, one can determine the source size independently from self-calibration, as shown in various previous publications for 3\,mm VLBI observations of Sgr\,A$^*$\xspace \citep{Doeleman2001, Bower2004, Shen2005, Bower2014, Ortiz2016}.\\

\noindent
In the context of this work, we are interested in a way to establish the observed size and orientation of Sgr\,A$^*$\xspace separately from self-cal. We therefore fit a simple model of an elliptical Gaussian component to the closure amplitude data, and we deconvolve the scattering ellipse using the best available model (Bower et al. 2006, 2014b) afterwards. We perform a $\chi^2$-analysis in fitting the Gaussian parameters (major and minor axis, and position angle).\\

\noindent
For both datasets, BF114A and BF114B, we derived the closure amplitudes from the 10s-averaged visibilities and fitted a simple 2D Gaussian source model to the closure amplitude data. There are some subtleties to take into account when modelfitting with closure amplitudes. $\chi^2$-minimization algorithms for model fitting generally assume that the errors on the measurements used are Gaussian. Closure amplitudes, when derived from visibilities with Gaussian errors, in general have non-Gaussian errors that introduce a potential bias when modelfitting which depends on the S/N and the relative amplitudes of the visibility measurements involved: because closure amplitudes are formed from a non-linear combination of visibility amplitudes (by multiplications and divisions), their error distribution is skewed (asymmetric). This is especially a problem in the low-S/N regime - the skew is much less pronounced for higher S/N values, and closure amplitude errors tend toward a Gaussian distribution in the high-S/N limit. Taking the logarithm of the measured closure amplitude values and appropriately defining the measurement uncertainties symmetrizes these errors, and generally results in more stable fitting results \citep{Chael2018}. For this reason, we adopt the technique described in that paper here.\\

\begin{figure}
\includegraphics[width=\textwidth]{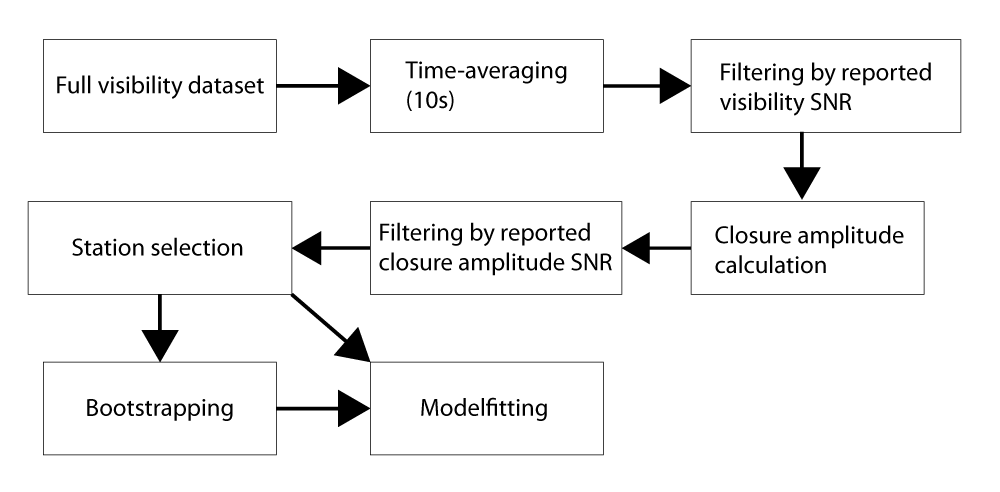}
 \caption{Overview of the pipeline used for closure amplitude model fitting. The stages involving time averaging, Visibility S/N filtering, Closure amplitude S/N filtering, station selection, and bootstrapping all offer different choices as to the parameters involved.}
\label{fig:flowchart-modelfitting}
\end{figure}

\noindent
The workflow we have adopted for the closure amplitude model fitting pipeline is outlined in Figure \ref{fig:flowchart-modelfitting}. We give a brief summary of the process here, and specify more details on individual steps below. We start the process with the frequency-averaged visibility dataset output from \textit{AIPS}, in which aberrant visibilites have already been flagged. We time-average this dataset to 10-second length segments using \textit{Difmap} to improve S/N per visibility measurement. In this step, the uncertainties on the resulting visibilities are recalculated using the scatter within each averaging period. The time interval of 10 seconds was experimentally confirmed to yield vector-averaged visibility amplitudes that are not significantly lower than when averaging over shorter timescales, and as such falls within the coherence timescale of the atmosphere at 86\,GHz. We also debias the averaged visibilities here, according to expression 9 in \citet{Chael2018}. We apply an S/N cutoff to the averaged 10 second visibility amplitudes at this point, where we have used different values for this cutoff to test the robustness of the model fitting results (described below). Using the remaining visibilities, we calculate the closure amplitudes for each 10 second time interval in the dataset. We calculate the error on these closure amplitude measurements using standard error propagation (following expression 12 from \citet{Chael2018}), and we then make another cut in the dataset where we discard all measurements that have a reported S/N below our threshold value. Lastly, we apply our station selection to the resulting dataset, dropping all closure amplitude measurements in which the omitted stations are involved. We thus obtain the dataset on which we perform model fitting.\\

\noindent
We use bootstrapping of the closure amplitudes of each dataset to determine the error on the individual fit parameters. Bootstrapping works by forming a new realization of measurement data by picking measurements from the original dataset at random (with replacement) until a new dataset is formed that has an equal number of measurements as the original dataset. As such, any measurement from the original dataset may be represented either once, multiple times or not at all in the newly formed dataset -- the weights of measurements in the original dataset are thus stochastically varied, emulating the drawing of a new sample of measurements. We fit the data with a 2D Gaussian model with three free parameters: major axis size, minor axis size and position angle on the sky of the major axis. The $\chi^2$ minimization is done as per expression 21 in \citet{Chael2018}.\\

\begin{figure*}
\centering
\includegraphics[width=0.32\textwidth]{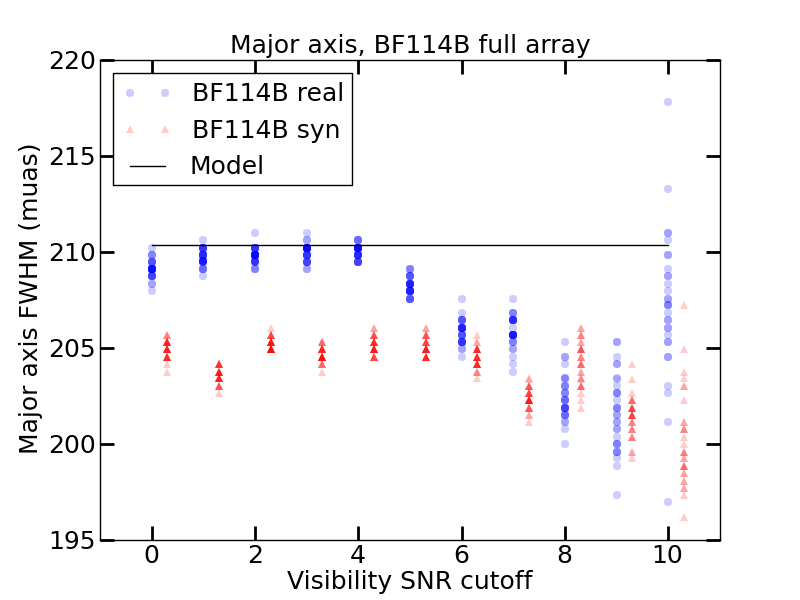}
\includegraphics[width=0.32\textwidth]{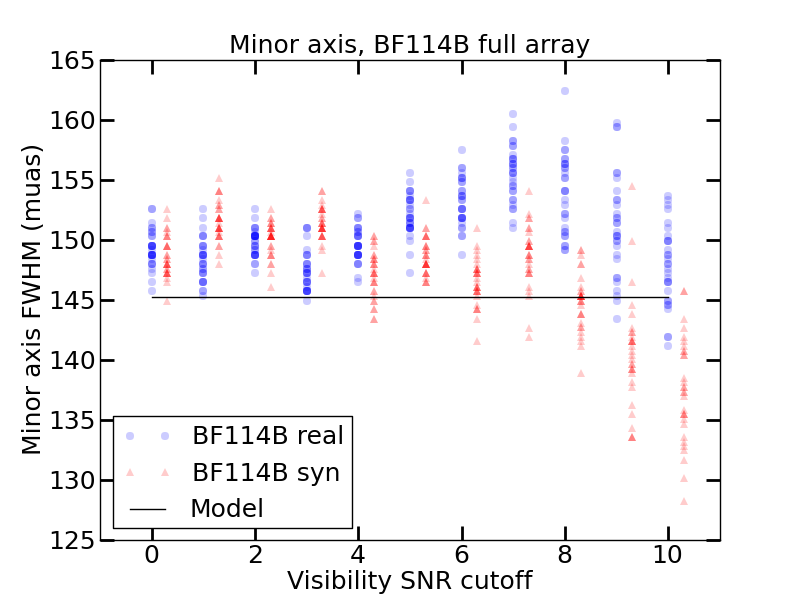}
\includegraphics[width=0.32\textwidth]{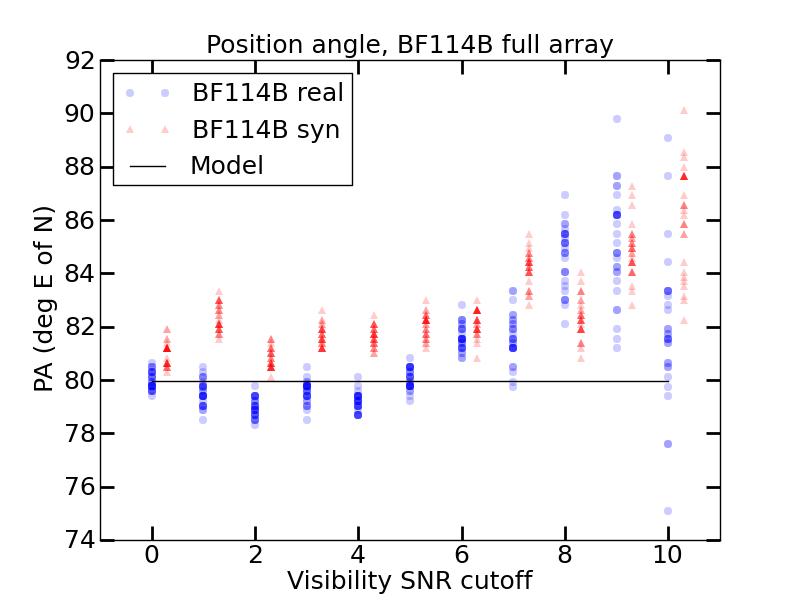}
\includegraphics[width=0.32\textwidth]{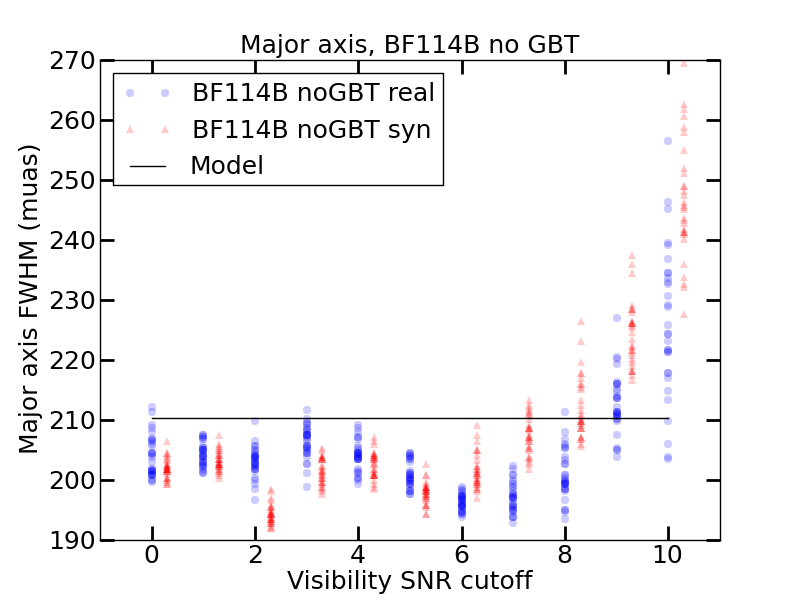}
\includegraphics[width=0.32\textwidth]{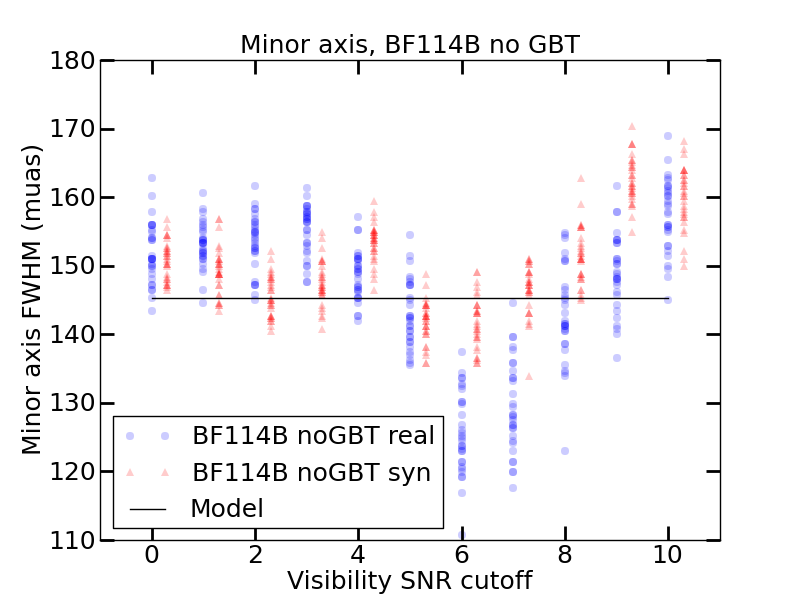}
\includegraphics[width=0.32\textwidth]{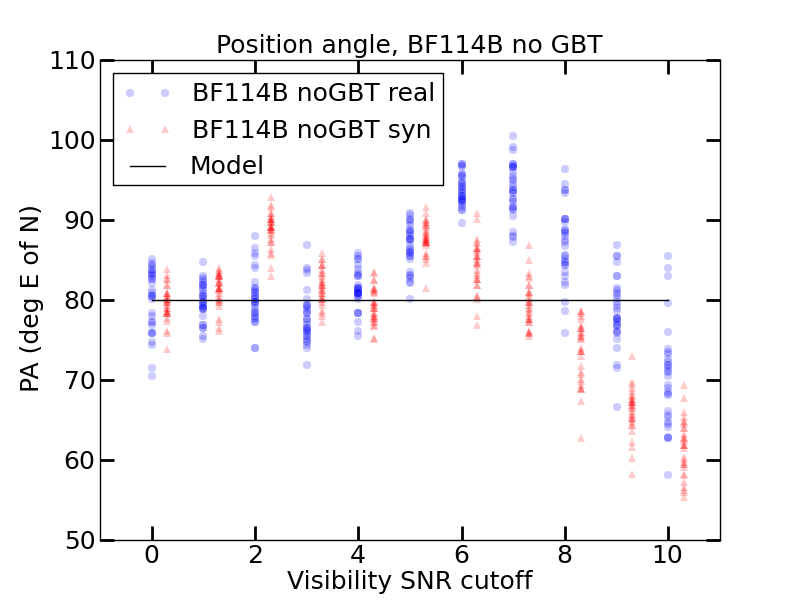}
\includegraphics[width=0.32\textwidth]{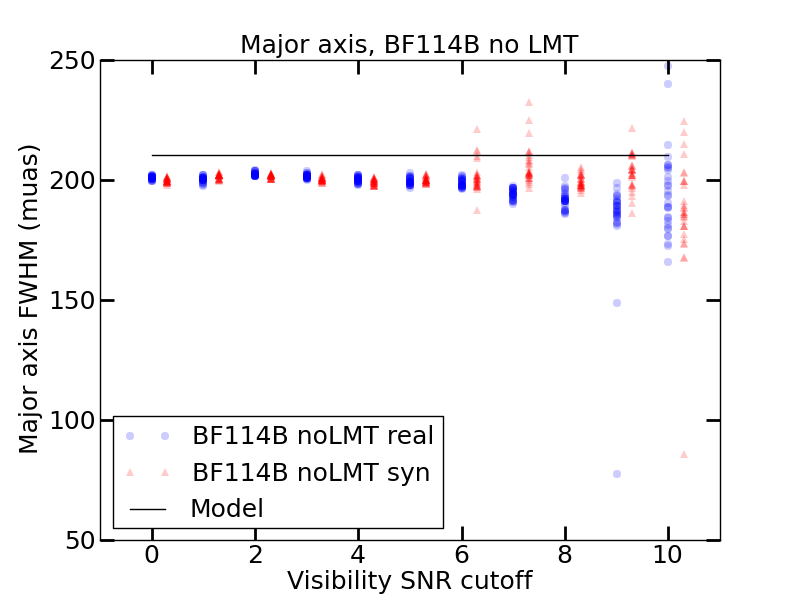}
\includegraphics[width=0.32\textwidth]{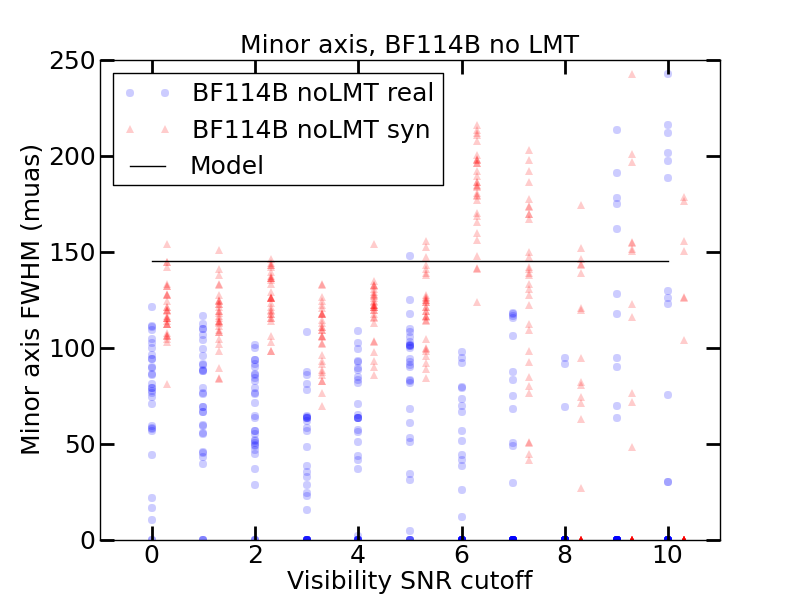}
\includegraphics[width=0.32\textwidth]{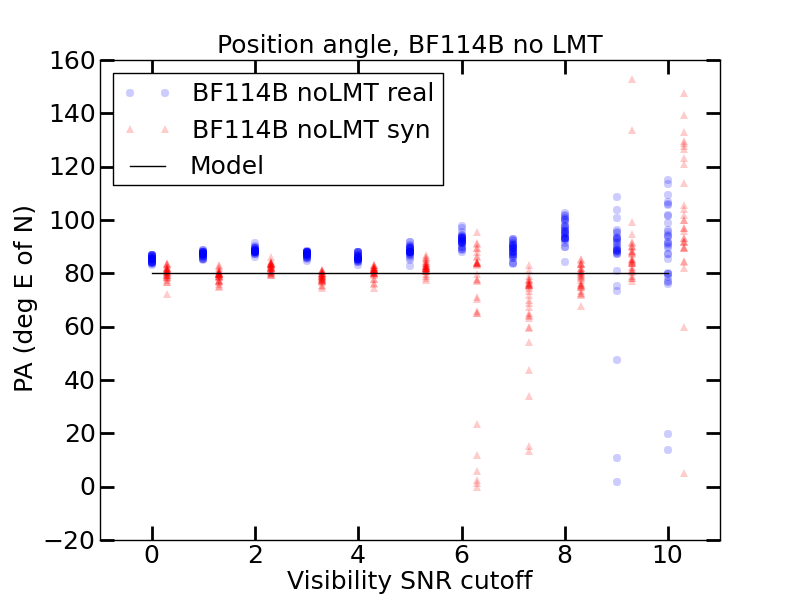}
\caption{Raw model fitting results for the BF114B dataset and for the synthetic dataset with the same $(u,v)$-sampling, using different integral S/N cutoff values and different station selections. The fiducial model parameters used to generate the synthetic dataset with are indicated by the horizontal black lines. For each S/N cutoff value, 31 bootstrapping realizations were performed to obtain uncertainties on the fitted model parameter values. Each of the results from these realizations is plotted with a single symbol. The different columns of figures show, from left to right, the major axis, minor axis and position angle results respectively. {\it Top row:}Full array, {\it Middle row:} without the GBT, {\it Bottom row:} without the LMT.}
\label{modelfitsBraw}
\end{figure*}

\noindent
Besides bootstrapping, we explore the effects of different values chosen for the S/N cutoff of the visibility amplitudes used in the model fitting. Visibilities with a low reported S/N are expected to have a larger influence on the skewness of the closure amplitude distribution, and are thus likely to introduce a bias in the fitting results. This effect is investigated by looking at different cutoff values for the visibility S/Ns. All visibility measurements can be assigned a 'reported S/N', which is defined as the measured visibility amplitude divided by the visibility amplitude uncertainty as determined from scatter among the measurements over a 10 second integration period. Before forming closure amplitudes using a visibility dataset, this visibility dataset is filtered by only admitting measurements that have reported S/Ns above a chosen threshold value. The constructed closure amplitudes can then be filtered again by their reported S/N. A closure amplitude S/N cutoff value of 3 was employed to avoid the larger bias that comes with low-S/N measurements, although we found that varying this value did not significantly impact the fitting results. The variation of visibility S/N cutoff has a more pronounced influence on fitting results, and this effect is shown in Figure \ref{modelfitsBraw}. The plots in the top row of this figure show the model fitting results for the full dataset, with all stations included. In these plots, where the blue circles indicate fitting results from the measured data, we see that the fitted model parameters show relatively minor variation over a range of S/N cutoff values from 1 to 4, where the minor axis size is the parameter that shows the largest spread. Above visibility S/N cutoff values of 4, we see that the spread in the fitting results grows and that trends of fitted values with S/N cutoff start appearing. This effect is coupled to the fact that only a limited number of quadrangles are left at these high S/N cutoff values, which by themselves provide weaker constraints on source geometry because of the limited $(u,v)$-coverage they provide.\\

\noindent
To investigate the consistency of the data regarding the convergence of best-fit model parameters, we also have performed model fits where we excluded the GBT from the array before gathering closure amplitude measurements and model-fitting. This was done to check if the inclusion of the GBT resulted in a systematic offset of fitted model parameters versus the case where the array does not include the GBT. Inclusion of the GBT offers a much better East-West array resolution, which is expected to have an impact on the quality of the major axis size estimate as the observed Sgr\,A$^*$\xspace Gaussian is oriented almost East-West on the sky. Likewise, the LMT offers a significant enhancement of the North-South array resolution and should therefore yield a clear improvement in quality for the estimated minor axis size. The model fitting results for these cases are included in Figure \ref{modelfitsBraw}, in the second (no GBT) and third (no LMT) rows. It is clear that indeed, inclusion of the GBT improves the quality of the major axis size estimate (the scatter among different bootstrapping realizations is significantly smaller than for the case where the GBT is omitted), while the LMT is instrumental in obtaining a good estimate for the minor axis size. As a result, the accuracy with which the position angle is determined benefits from inclusion of both the GBT and the LMT.

\begin{figure*}
\centering
\includegraphics[width=0.48\textwidth]{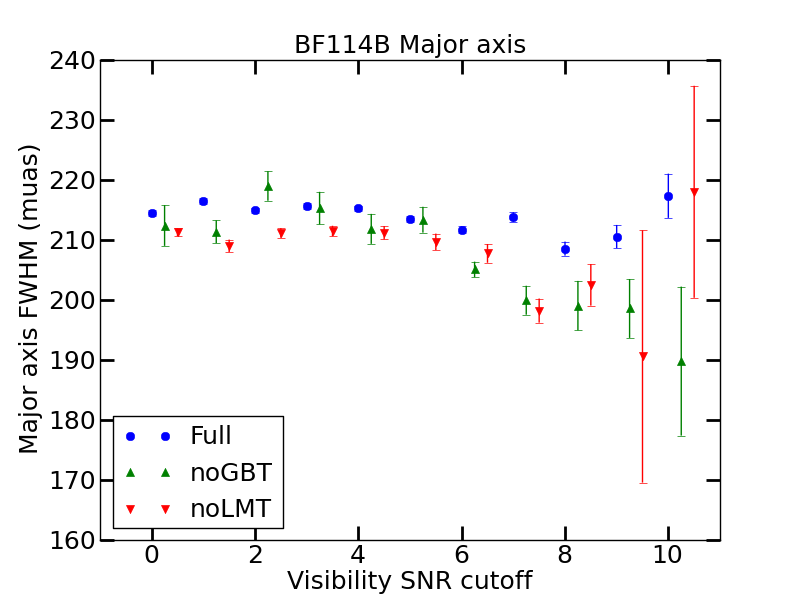}
\includegraphics[width=0.48\textwidth]{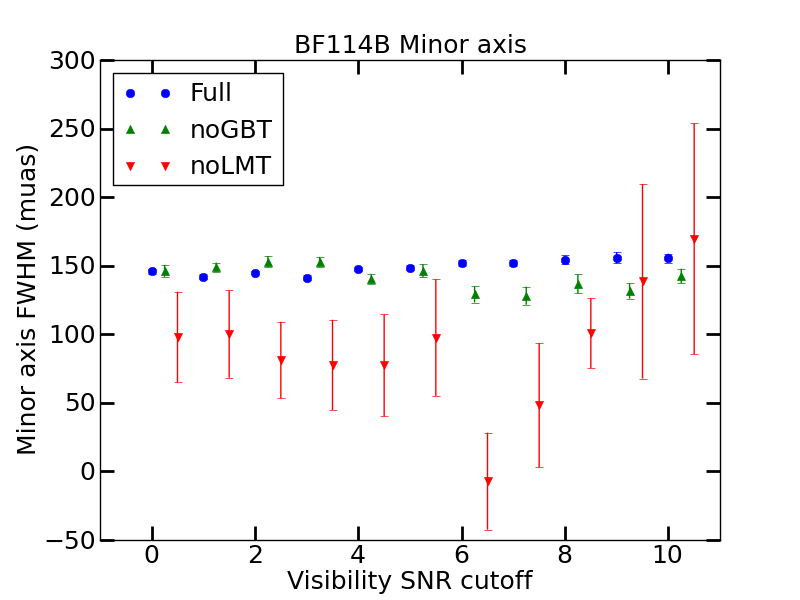}
\includegraphics[width=0.48\textwidth]{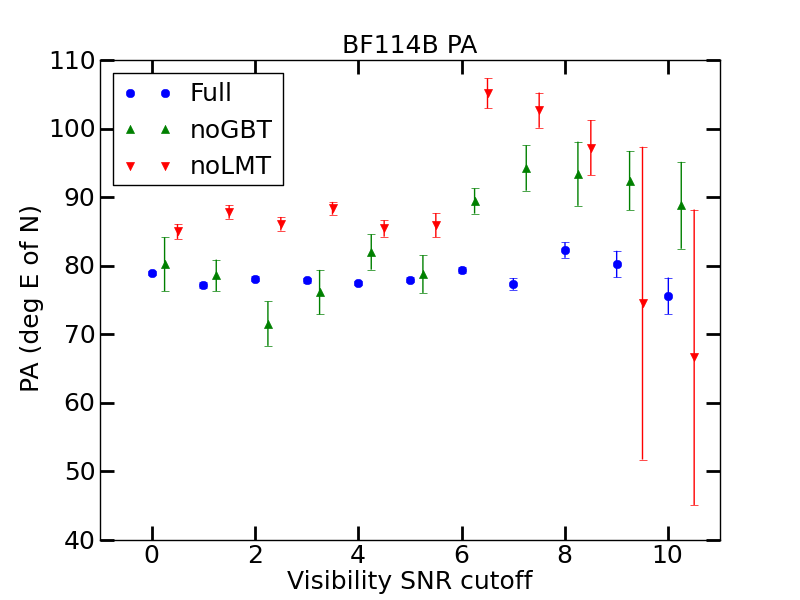}
\includegraphics[width=0.48\textwidth]{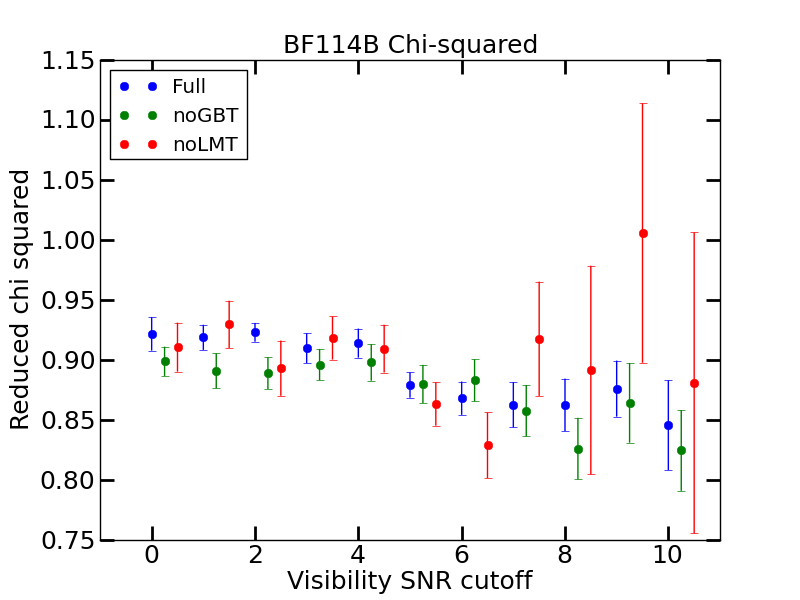}
\caption{Bias-corrected model fitting results for the BF114B dataset for different station selections as a function of visibility S/N cutoff value. The fitted parameter values for the measured data have been corrected using the offset exhibited by the fits to the synthetic datasets. The results per station selection (symbol type) have been offset along the S/N axis by a small amount for clarity.}
\label{modelfitsB}
\end{figure*}

\noindent
We should note that consistency of fitted model parameters by itself does not guarantee accurate results (only precise results). For this reason, we have generated synthetic visibility datasets with the same $(u,v)$-sampling as the original measurements, where a Gaussian source model with fiducial parameter values that are close to the previously measured size of Sgr\,A$^*$\xspace (Major axis: 210.4 $\mu$as, minor axis: 145.2 $\mu$as, position angle: 80 degrees East of North) was used as input. The visibility uncertainties for this synthetic dataset were scaled in such a way as to yield the same distribution in S/N values as the original data shows. For this synthetic dataset, the full processing pipeline was then used and the deviations of the fitted parameters from the fiducial inputs were inspected. These results are also plotted in Figure \ref{modelfitsBraw}, using red triangles as markers for the model fitting results and black lines to indicate the input model parameter values. For the major axis size, we see that the fitted values typically underpredict the actual source size by 5 to 10 $\mu$as, depending on which stations are involved in the array. The minor axis size is severely underpredicted when the LMT is left out of the array, but is close to the input value when the LMT is included. The position angle come out close to the input value in all cases, although there is a small positive bias seen in the case where the full array is used. Note that the y-axis ranges of these plots are different, and that the spread seen in the case of the full array are typically much smaller than those for the other array configurations. These results from synthetic data fitting allow us to correct for the biases that our pipeline exhibits. The bias-corrected fitted source parameters are shown in Figure \ref{modelfitsB}. For all model parameters we get consistent fitting results for all visibility S/N cutoff choices up to 5. Because any specific choice of S/N cutoff value is difficult to defend for coming up with our final model parameter fitting values, we note that the scatter of the fitted values among these different visibility S/N cutoffs is consistent with their uncertainties in most cases. We therefore use the average value for the model fit results up to and including the S/N cutoff of 5, and for the uncertainty we use the average uncertainty for the same data points. Our derived source geometry parameters are listed in Table \ref{table:sizes}, together with previously reported sizes. We intentionally do not quote estimates of the intrinsic size for Sgr\,A$^*$\xspace, for reasons explained in the following section.

\begin{table*}
\centering
\scriptsize
\addtolength{\leftskip} {-2cm}
\addtolength{\rightskip}{-2cm}
\caption{Sgr~A*: size of elliptical Gaussian fits to observed 86\,GHz emission}
\label{table:sizes}
\begin{tabular}{ l |  c c c  c | c}
\hline\hline
Reference  & major axis & minor axis & position angle & axial ratio & calculated intrinsic size\\
                 & [$\mu$as]   & [$\mu$as]    & [$^\circ$]        &[-]                  &[$\mu$as]\\
\hline 
\citet{Ortiz2016} (obs. 1, self-cal.) & $212.7\pm2.3$ & $138.5\pm3.5$ & $81.1\pm1.8$ & $1.54\pm0.04$ & $142\pm9 \times 114\pm15$\\
\citet{Ortiz2016} (obs. 2,  self-cal.)   & $221.7\pm3.6$ & $145.6\pm4.0$ & $75.2\pm2.5$ & $1.52\pm0.05$ &$155\pm9\times122\pm14$\\
\citet{Lu2011_sgra} (self-cal) & $210\pm10$ & $130\pm10$ & $83.2\pm1.5$ &1.62 & $139\pm17\times102\pm21$\\
\citep{Shen2005} (clos. ampl.)& $210^{+20}_{-10}$ & $130^{+50}_{-13}$ & $79^{+12}_{-33}$ & 1.62 \\
\citep{Doeleman2001} (self-cal., averaged) & $180\pm20$ & -- & -- & -- \\
\citep{Krichbaum1998} (modelfit) & $190\pm30$ & -- & -- & -- \\
\hline
this work (self-cal) & $217\pm22$ & $165\pm17$ & $77\pm15$ & 1.3 & $167\pm22 \times 122\pm25^{*}$\\
this work (clos. ampl., full array) & $215.1\pm0.4$ & $145.1\pm1.5$ & $77.9\pm0.4$ & $1.48\pm0.01$ & $145.4\pm0.6 \times 122.6\pm1.7$\\
this work (clos. ampl., no GBT) & $213.9\pm2.5$ & $148.0\pm4.0$ & $77.9\pm3.0$ & $1.45\pm0.03$ & $144.4\pm3.7 \times 125.2\pm4.9$\\
this work (clos. ampl., no LMT) & $210.6\pm1.0$ & $88.7\pm34.2$ & $86.4\pm1.2$ & $2.37\pm1.02$ & $86.5\pm69.7 \times 40.6\pm40.5$\\
\hline
\end{tabular}
\newline
$^{*}$Calculated using a scattering kernel size of $158.5 \times 77.5$ $\mu$as at 86\,GHz, from \citet{Bower2006}. No uncertainty in scattering kernel size was incorporated in this calculation. Intrinsic sizes from our closure amplitude results in this table also use this scattering kernel.
\end{table*}

\section{Constraints on the size-frequency relation and the scattering law}

\noindent
Extensive measurements of the size of Sgr\,A* have been performed over the years at various frequencies, leading to an understanding of the nature of the scattering law in the direction of the Galactic center \citep{Backer1978, Lo1998, Bower2006, Johnson2015, Psaltis2015} as well as on the dependency of intrinsic source size on frequency both from an observational and a theoretical perspective \citep{Bower2004, Bower2006, Shen2006, Bower2014, Moscibrodzka2014, Ortiz2016}. Knowledge of the intrinsic source size at different frequencies is an important component of the research on Sgr\,A*, because it strongly constrains possible models for electron temperatures, jet activity and particle acceleration.\\

\noindent
Our size measurements of Sgr\,A* at 86\,GHz, when combined with these previously published size measurements over a range of frequencies, allow us to perform a simultaneous fitting of the size-frequency relation together with the scattering law. Previous studies have focused on constraining either the scattering law or the intrinsic size-frequency relation, typically by either focusing on a specific range of longer observing wavelengths to constrain the scattering law \citep{Psaltis2015} or by using a fiducial scattering law and focusing on the shorter observing wavelengths to establish an intrinsic size-frequency relation \citep{Bower2006}. However, simultaneous fits of both of these relations to the available data have not been published to date. \citet{Johnson2018} find a size $b_{maj}=1.380 \pm 0.013 \left({\lambda \over \textrm{cm}}\right)^2$ milliarcseconds using a similar set of past results and analysis techniques as used in this work. The difference with our constraint emphasizes the challenge of obtaining a solution with 1\% precision in the complex domain of heterogenous data sets, extended source structure, and an unknown intrinsic size.\\

\noindent
Besides our measurements presented in this paper, we use previously published size measurements from \citet{BowerBacker1998, Krichbaum1998, Bower2004, Bower2006, Shen2006, Doeleman2008, Bower2014, Ortiz2016}, where \citet{Bower2004} includes re-analysed measurements originally published in \citet{Lo1998}. Care was taken to ensure that all these published results were derived from data that was independently obtained and analyzed. The measurements we include for the model fitting have been taken over a time period of multiple decades, thereby most likely representing different states of activity of the source which may affect size measurements. This effect is expected to be small, however: at short wavelengths because of the stable source size that has been measured over time, and at longer wavelengths because the scattering size is so much larger than the intrinsic size. The measurements taken at wavelengths close to $\lambda=20$ cm were taken closely spaced in time, yet still show a mutual scatter that is wider than the size of their error bars suggests: this may indicate the presence of systematics in the data. An ongoing re-analysis of these sizes at long wavelengths \citep{Johnson2018} suggests that these measurements are too small by up to 10\%, likely impacting the resulting fits for the scattering law and intrinsic size-frequency relation. Here, we use the values as they have been published. Throughout this section, we use Gaussian models for both the observed source size and for the scattering kernel. Recent work has shown that the instantaneous shape of the scattering kernel deviates from a Gaussian to a limited extent \citep{Gwinn2014_scatteringsubstructure}, but the statistical average of the scattering kernel geometry is thought to be Gaussian to within a few percent.\\

\begin{figure*}
\centering
\includegraphics[width=0.53\textwidth]{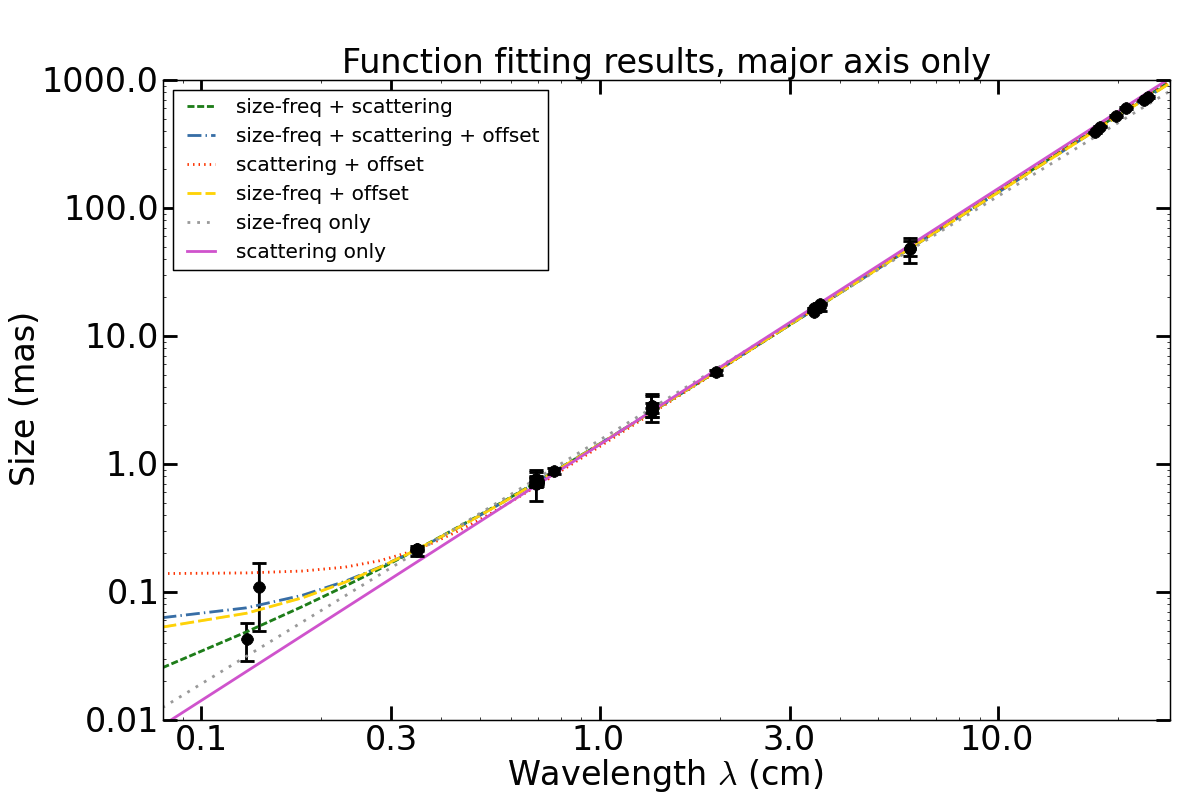}
\includegraphics[width=0.46\textwidth]{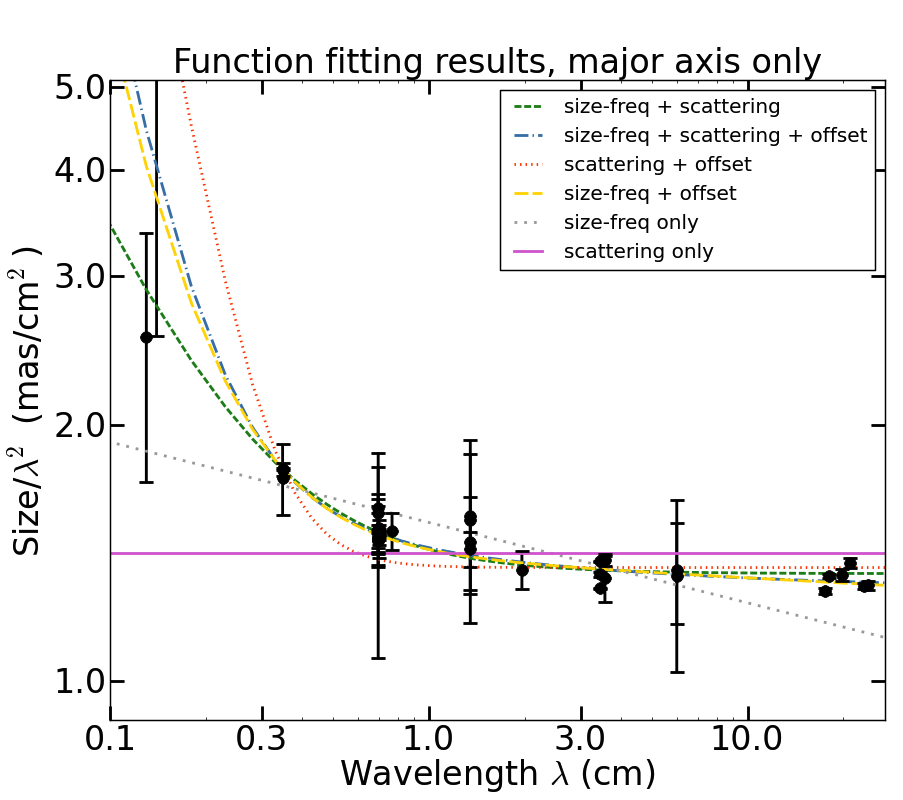}
\caption{Left: aggregate measurement data for the observed major axis size of Sgr\,A* (black points with error bars), and model fitting results for different combinations of included model components (coloured lines). The highest-quality fits are provided by the green, blue and orange lines (the top 3 listed in the legend) which provide very similar fit qualities (see Table \ref{tab:modelfits}). Right: the same data, plotted with the major axis sizes divided by wavelength squared. The fitting results without the 230\,GHz data are almost identical to these, and hence are not plotted separately.}
\label{fig:sizefreqfitting4}
\end{figure*}

\noindent
The set of measurements, as we have used them in the model fitting, are visible in Figure \ref{fig:sizefreqfitting4}. Measurements taken at the highest of these frequencies (230\,GHz) are expected to feature emission coming from very close to the black hole shadow, and as such the perceived source size may be significantly affected by gravitational lensing effects where the source image can be warped into a crescent-like structure. Such strong lensing effects are not expected to play a role in source sizes as observed at lower frequencies because the inner accretion flow is optically thick at small radii for those frequencies. We thus expect to effectively see emission coming from somewhat larger radii where the light paths are not significantly affected by spacetime curvature but are affected by interstellar scattering along our line of sight. We have therefore done the model fitting both including the 230\,GHz size measurements (Figure \ref{fig:sizefreqfitting4}) and excluding them, to see if the expected GR lensing effects play a significant role in the appearance of the source at the shortest wavelengths. We find very little difference in the best-fit parameter values between the results.\\

\noindent
Simultaneous fitting of the size-frequency relation and the scattering law is done using the major axis size measurements only, as the uncertainties in the minor axis size measurements are too large to provide any meaningful constraint on the models. For the size-frequency relation, we use the following expression:\\

\begin{equation}
\theta_{\textrm{int}}(\lambda)=a \cdot \lambda^{b},
\end{equation}

\noindent
where $a$ and $b$ are constants to be determined, $\theta_{\textrm{int}}$ is the intrinsic angular size in milliarcseconds and $\lambda$ is the observing wavelength in cm. For the scattering law we adopt the expression:

\begin{equation}
\theta_{\textrm{scatt}}(\lambda)=c \cdot \lambda^{2},
\end{equation}
\noindent
where $c$ is a constant to be determined and with $\theta_{\textrm{scatt}}$ the angular broadening through scattering in milliarcseconds. These sizes are added in quadrature to provide the measured major axis size for Sgr\,A*:

\begin{equation}
\theta_{\textrm{meas}}(\lambda)=\sqrt{\theta^{2}_{\textrm{int}} + \theta^{2}_{\textrm{scatt}}}.
\end{equation}

\noindent
This expression is used in the fitting procedure to obtain a measured size from the model parameters, thus involving at most 3 free parameters (the constants $a$, $b$, and $c$). Using a simple linear least-squares fitting procedure (from the Python package scipy.optimize.curve\_fit), and fitting to all size measurement data available, we get the following values and uncertainties in the expressions for intrinsic size and scattering size respectively (see also Figure \ref{fig:sizefreqfitting4} for the model curves produced):\\

\begin{equation}
\theta_{\textrm{int}}(\lambda) = 0.502\pm0.075 \cdot \lambda^{1.201\pm0.138},
\label{sf230}
\end{equation}

\begin{equation}
\theta_{\textrm{scatt}}(\lambda) = 1.338\pm0.012 \cdot \lambda^{2}.
\label{scat230}
\end{equation}

\noindent
At 230\,GHz, there is the possibility that the size of Sgr\,A$^*$\xspace may be strongly affected by gravitational lensing. To investigate whether the inclusion of these measurements significantly affects the size/wavelength relation found, we also perform the fitting routine while leaving out the 230\,GHz measurements. We then get the following expressions for intrinsic size and scattering size:\\

\begin{equation}
\theta_{\textrm{int}}(\lambda) = 0.502\pm0.078 \cdot \lambda^{1.201\pm0.143},
\label{sfno230}
\end{equation}

\begin{equation}
\theta_{\textrm{scatt}}(\lambda) = 1.338\pm0.012 \cdot \lambda^{2}.
\label{scatno230}
\end{equation}

\noindent
Cross-comparing expressions \ref{sfno230} and \ref{scatno230} to \ref{sf230} and \ref{scat230}, we see that the corresponding fitted model parameters between the model fits with and without the 230\,GHz measurements are well within each other's error bars for all 3 model parameters. The available measurements of source size at 1.3\,mm thus seem to be compatible with the source size as predicted using the fitted size/wavelengths relations from the other measurements.\\

\noindent
Comparing these figures to \citet{Bower2015pulsar}, we see that the scattering size parameter for the major axis is well within the error bars of the value calculated in that work ($b_{\textrm{maj, scatt}} = 1.32 \pm 0.02$ mas cm$^{-2}$).  For the intrinsic size as a function of wavelength, the powerlaw index we find is somewhat larger than the powerlaw index calculated in \citet{Ortiz2016} (where it is quoted as being $1.34 \pm 0.13$), but still within the error bars.\\

\noindent
The size/wavelength relation that we have used up to this point has a specific functional form: it consists of a pure powerlaw for the size/frequency relation, combined in quadrature with a scattering law where scattering size scales with wavelength squared. To explore the influence that this choice of functional form has on the results of the fitting procedure, we have performed the fit with other models for the dependence of observed size on observing wavelength as well. All models consist of a combination of three components: a fixed-size component that is constant across all wavelengths, a scaled $\lambda^p$ component (where $p$ is a free parameter) that is linearly added to it, and a scaled $\lambda^2$ component (scattering law) that is then added to the sum of the other component(s) in quadrature. Six combinations of these model components were fitted to the major axis size measurement data, and each fit was done for two cases: with and without the 230 GHz observed source sizes included in the data to be fitted to. In Table \ref{table:sizemodels}, the results of these model parameter fits are presented.\\

\noindent
The three best-fitting models are the 'regular' model (scattering law + general power law), the 'augmented' model (scattering law + general power law + fixed size offset), and the 'simple' model (general power law + fixed size offset). For the 'simple' model, the best-fitting power law index is close to 2 within a few percent. If the power law exponent from scattering can deviate from the theoretically ideal value of 2 by even a small fraction, this result suggests that the intrinsic size-frequency relation for Sgr\,A$^*$\xspace is less certain than what has been found in previous publications. A similar conclusion was derived by \citet{Bower2006}, where it was found that a relaxation of the scattering exponent to values slightly different from 2 undercuts the support for an intrinsic size/frequency relation with a nonzero power law index.\\

\begin{table*}
\centering
\scriptsize
\addtolength{\leftskip} {-2cm}
\addtolength{\rightskip}{-2cm}
\caption{Sgr~A*: fitted size dependence on frequency, different models}\label{table:sizemodels}
\begin{tabularx}{\textwidth}{ p{35mm} | p{7mm} c c c c c c}
\hline\hline
Model  & incl. 230 GHz? & a & b & c & d & $\chi^2$ / d.o.f.\\
\hline 
\hline
Size-freq + scattering \newline $\sqrt{(a\lambda^2)^2 + (b\lambda^c)^2}$ \vspace{2mm} & yes & 1.338  $\pm$  0.012  &  0.502  $\pm$  0.075  &  1.201  $\pm$  0.138  & - &  1146.29  /  34 \\
Size-freq + scattering + offset \newline $\sqrt{(a\lambda^2)^2 + (b\lambda^c + d)^2}$ \vspace{2mm} & yes &  1.277  $\pm$  0.110  &  0.600  $\pm$  0.205  &  1.757  $\pm$  0.320  &  0.055  $\pm$  0.021  &  1107.79  /  33 \\
Scattering + offset \newline $\sqrt{(a\lambda^2)^2 + d^2}$ \vspace{2mm} & yes &  1.360  $\pm$  0.009  & - & - &  0.139  $\pm$  0.005  &  1873.88  /  35 \\
Size-freq + offset \newline $b\lambda^c + d$ \vspace{2mm} & yes &   - &  1.385  $\pm$  0.019  &  1.980  $\pm$  0.010  &  0.044  $\pm$  0.005  &  1108.29  /  34 \\
Size-freq only \newline $b\lambda^c$ \vspace{2mm} & yes &   - &  1.537  $\pm$  0.015  &  1.905  $\pm$  0.008  & - &  3292.23  /  35 \\
Scattering only \newline $a\lambda^2$ \vspace{2mm} & yes &  1.417  $\pm$  0.024  & - & - & - &  15944.28  /  36 \\
\hline
$\sqrt{(a\lambda^2 )^2+ (b\lambda^c)^2}$ & no &  1.338  $\pm$  0.012  &  0.502  $\pm$  0.078  &  1.201  $\pm$  0.143  & - &  1145.25  /  32 \\
$\sqrt{(a\lambda^2)^2 + (b\lambda^c + d)^2}$ & no &  1.273  $\pm$  0.128  &  0.606  $\pm$  0.235  &  1.773  $\pm$  0.337  &  0.057  $\pm$  0.021  &  1102.07  /  31 \\
$\sqrt{(a\lambda^2)^2 + d^2}$ & no &  1.360  $\pm$  0.009  & - & - &  0.139  $\pm$  0.005  &  1824.55  /  33 \\
$b\lambda^c + d$ & no &   - &  1.385  $\pm$  0.020  &  1.980  $\pm$  0.010  &  0.044  $\pm$  0.005  &  1104.63  /  32 \\
$b\lambda^c$ & no &   - &  1.537  $\pm$  0.015  &  1.905  $\pm$  0.008  & - &  3290.05  /  33 \\
$a\lambda^2$ & no &  1.417  $\pm$  0.025  & - & - & - &  15940.55  /  34 \\
\end{tabularx}
\label{tab:modelfits}
\end{table*}

\section{Summary and Conclusions}\label{sec:summary}

Constraining the intrinsic size and structure of Sgr\,A$^*$\xspace at an observing wavelength of 3\,mm still remains a challenge. Although the effect of interstellar scattering becomes smaller at this wavelength, it is still not negligible. GRMHD models of the accretion flow around Sgr\,A$^*$\xspace \citep[e.g.,][]{Moscibrodzka2014} predict a certain structure in the emission which should be detectable with current VLBI arrays. However, detection of intrinsic substructure could be hindered by refractive scattering, possibly itself introducing compact emission substructure \citep{JohnsonGwinn2015}.\\

\noindent
In this paper, we have presented imaging results and analysis of closure amplitudes of new VLBI observations performed with the VLBA, the LMT and the GBT at 86\,GHz. Following our previous result (Paper~I) from the analysis of closure phases, the detection of substructure in the 3\,mm emission of Sgr\,A$^*$\xspace, we confirm the previous result of compact substructure using imaging techniques. Using NRAO\,530\xspace as test source, we show that VLBI amplitude calibration can be performed with an absolute uncertainty of 20\% for NRAO\,530\xspace and 30\% for Sgr\,A$^*$\xspace, where we are currently limited by the uncertainty in antenna gains. The variable component of these gain uncertainties is limited to $\sim$10\%.\\

\noindent
Out of our two experiments, only in the higher resolution and more sensitive experiment (BF114B, including the VLBA, the LMT and the GBT) is the compact asymmetric emission clearly detected. The VLBA+LMT dataset (BF114A) remains inconclusive in this respect. The asymmetry is detected as significant residual emission, when modeling the emission with an elliptical Gaussian component. The flux density of the asymmetrical component is about 10\,mJy. Such a feature can be explained by refractive scattering, which is expected to result in an RMS flux of this level, but an intrinsic origin cannot be excluded. The discrimination and disentanglement of both these possible origins requires a series of high-resolution and multifrequency VLBI observations, spread out in time. Interestingly, the secondary off-core component observed at 7\,mm with the VLBA \citep{Rauch2016} is found at a similar position angle. The authors of that paper interpret this feature as an adiabatically expanding jet feature. Future, preferably simultaneous, 3\,and 7\,mm VLBI observations can shed light on the specific nature of the compact emission. A persistent asymmetry, observed over multiple epochs that are spaced apart in time by more than the scattering timescale at 86\,GHz, would provide strong evidence for an intrinsic source asymmetry. Another way in which observed asymmetry may be ascribed to source behaviour rather than scattering is when a transient asymmetry evolution is accompanied by a correlated variation in integrated source flux density. Observations of that nature will require succesive epochs using a consistent and long-baseline array of stations involved accompanied by independent high-quality integrated flux density measurements (e.g., by ALMA).\\

\noindent
We see that the combination of the VLBA, LMT and GBT provides the capability to pin down the observed source geometry with unsurpassed precision because of the combination of sensitivity and extensive $(u,v)$-coverage provided, going beyond what addition of the LMT or the GBT separately can do. This combination of facilities is therefore important to involve in future observations that aim to measure the geometry of Sgr\,A$^*$\xspace.\\

\noindent
We also note that even with this extended array, the measurement and characterization of complex source structure beyond a 2D Gaussian source model is something that remains difficult. To study Sgr\,A$^*$\xspace source substructure at 86\,GHz more closely, be it either intrinsic or from scattering, even more extensive $(u,v)$-coverage and sensitivity will be needed. Recent measurements done with GMVA + ALMA, the analysis of which is underway, should allow for a more advanced study of the complex source structure of Sgr\,A$^*$\xspace, as that array configuration provides unprecedented North-South $(u,v)$-coverage combined with high sensitivity on those long baselines.\\

\noindent
Moving from source sub-structure to overall geometry, this work has reported the observed source geometry of Sgr\,A$^*$\xspace with the highest accuracy to date. Addition of the GBT adds East-West resolving power as well as extra sensitivity and redundancy in terms of measured visibilities. We note that the source geometry we find is very similar to that reported in \citep{Ortiz2016}, while the different observations were spaced almost one month apart (April 27th for BD183C, May 23rd for BF114B). Barring an unlikely coincidence, this suggests a source geometry that is stable to within just a few percent over that time scale. At 86\,GHz, Sgr\,A$^*$\xspace is known to exhibit variability in amplitude at the $\sim$10\% level (see Paper~I) on intra-day timescales. Whether these short-timescale variations in flux density correspond to variations in source size is an open question that can only be resolved when dense $(u,v)$-coverage is available at high sensitivity (beyond current capabilities), as source size would need to be accurately measured multiple times within a single epoch. Alternatively, studies of the source size variability at somewhat longer timescales can simply be done by observing Sgr\,A$^*$\xspace over multiple epochs -- but the fast variations will be smeared out as a result.\\

\noindent
From the simultaneous fitting of the scattering law and the intrinsic size/frequency relation for Sgr\,A$^*$\xspace, we find values compatible with existing published results. However, if the scattering law is allowed to deviate from a pure $\lambda^2$ law toward even a slightly different power law index, differing by e.g. $2\%$ from the value 2, support for the published intrinsic size/frequency relation often used in the literature quickly disappears. We therefore advocate a cautious stance towards the weight given to existing models for the intrinsic size-frequency relation for Sgr\,A$^*$\xspace.\\

{\it Acknowledgements:} We thank the anonymous referee for providing comments that improved the quality of the paper. C.B. wishes to thank Michael Johnson and Lindy Blackburn for valuable discussions which improved the robustness of the closure amplitude analysis. This work is supported by the ERC Synergy Grant “BlackHoleCam: Imaging the Event Horizon of Black Holes”, Grant 610058, \cite{bhc_rev}. L.L. acknowledges the financial support of DGAPA, UNAM (project IN112417), and CONACyT, M\'exico. S.D. acknowledges support from National Science Foundation grants AST-1310896, AST-1337663 and AST-1440254.

\bibliographystyle{aa}
\bibliography{SGRA_NRAO530}

\begin{thebibliography}{44}
\expandafter\ifx\csname natexlab\endcsname\relax\def\natexlab#1{#1}\fi

\bibitem[{{An} {et~al.}(2013){An}, {Baan}, {Wang}, {Wang}, \& {Hong}}]{An2013}
{An}, T., {Baan}, W.~A., {Wang}, J.-Y., {Wang}, Y., \& {Hong}, X.-Y. 2013,
  MNRAS, 434, 3487

\bibitem[{{Backer}(1978)}]{Backer1978}
{Backer}, D.~C. 1978, ApJ L, 222, L9

\bibitem[{{Bower} \& {Backer}(1998)}]{BowerBacker1998}
{Bower}, G.~C. \& {Backer}, D.~C. 1998, ApJ L, 496, L97

\bibitem[{{Bower} {et~al.}(2014{\natexlab{a}}){Bower}, {Deller}, {Demorest},
  {Brunthaler}, {Eatough}, {Falcke}, {Kramer}, {Lee}, \&
  {Spitler}}]{Bower2014_pulsar}
{Bower}, G.~C., {Deller}, A., {Demorest}, P., {et~al.} 2014{\natexlab{a}},
  ApJ L, 780, L2

\bibitem[{{Bower} {et~al.}(2015){Bower}, {Deller}, {Demorest}, {Brunthaler},
  {Falcke}, {Moscibrodzka}, {O'Leary}, {Eatough}, {Kramer}, {Lee}, {Spitler},
  {Desvignes}, {Rushton}, {Doeleman}, \& {Reid}}]{Bower2015pulsar}
{Bower}, G.~C., {Deller}, A., {Demorest}, P., {et~al.} 2015, ApJ, 798, 120

\bibitem[{{Bower} {et~al.}(2004){Bower}, {Falcke}, {Herrnstein}, {Zhao},
  {Goss}, \& {Backer}}]{Bower2004}
{Bower}, G.~C., {Falcke}, H., {Herrnstein}, R.~M., {et~al.} 2004, Science, 304,
  704

\bibitem[{{Bower} {et~al.}(2006){Bower}, {Goss}, {Falcke}, {Backer}, \&
  {Lithwick}}]{Bower2006}
{Bower}, G.~C., {Goss}, W.~M., {Falcke}, H., {Backer}, D.~C., \& {Lithwick}, Y.
  2006, ApJ, 648, L127

\bibitem[{{Bower} {et~al.}(2014{\natexlab{b}}){Bower}, {Markoff}, {Brunthaler},
  {Law}, {Falcke}, {Maitra}, {Clavel}, {Goldwurm}, {Morris}, {Witzel}, {Meyer},
  \& {Ghez}}]{Bower2014}
{Bower}, G.~C., {Markoff}, S., {Brunthaler}, A., {et~al.} 2014{\natexlab{b}},
  ApJ, 790, 1

\bibitem[{{Brinkerink} {et~al.}(2016){Brinkerink}, {M{\"u}ller}, {Falcke},
  {Bower}, {Krichbaum}, {Castillo}, {Deller}, {Doeleman}, {Fraga-Encinas},
  {Goddi}, {Hern{\'a}ndez-G{\'o}mez}, {Hughes}, {Kramer}, {L{\'e}on-Tavares},
  {Loinard}, {Monta{\~n}a}, {Mo{\'s}cibrodzka}, {Ortiz-Le{\'o}n},
  {Sanchez-Arguelles}, {Tilanus}, {Wilson}, \& {Zensus}}]{Brinkerink2016}
{Brinkerink}, C.~D., {M{\"u}ller}, C., {Falcke}, H., {et~al.} 2016, MNRAS,
  462, 1382

\bibitem[{{Broderick} {et~al.}(2016){Broderick}, {Fish}, {Johnson},
  {Rosenfeld}, {Wang}, {Doeleman}, {Akiyama}, {Johannsen}, \&
  {Roy}}]{Broderick2016}
{Broderick}, A.~E., {Fish}, V.~L., {Johnson}, M.~D., {et~al.} 2016, ApJ, 820,
  137

\bibitem[{{Chael} {et~al.}(2018){Chael}, {Johnson}, {Bouman}, {Blackburn},
  {Akiyama}, \& {Narayan}}]{Chael2018}
{Chael}, A.~A., {Johnson}, M.~D., {Bouman}, K.~L., {et~al.} 2018, ApJ, 857, 23

\bibitem[{{Davies} {et~al.}(1976){Davies}, {Walsh}, \& {Booth}}]{Davies1976}
{Davies}, R.~D., {Walsh}, D., \& {Booth}, R.~S. 1976, MNRAS, 177, 319

\bibitem[{{Doeleman} {et~al.}(2001){Doeleman}, {Shen}, {Rogers}, {Bower},
  {Wright}, {Zhao}, {Backer}, {Crowley}, {Freund}, {Ho}, {Lo}, \&
  {Woody}}]{Doeleman2001}
{Doeleman}, S.~S., {Shen}, Z.-Q., {Rogers}, A.~E.~E., {et~al.} 2001, AJ, 121,
  2610

\bibitem[{{Doeleman} {et~al.}(2008){Doeleman}, {Weintroub}, {Rogers},
  {Plambeck}, {Freund}, {Tilanus}, {Friberg}, {Ziurys}, {Moran}, {Corey},
  {Young}, {Smythe}, {Titus}, {Marrone}, {Cappallo}, {Bock}, {Bower},
  {Chamberlin}, {Davis}, {Krichbaum}, {Lamb}, {Maness}, {Niell}, {Roy},
  {Strittmatter}, {Werthimer}, {Whitney}, \& {Woody}}]{Doeleman2008}
{Doeleman}, S.~S., {Weintroub}, J., {Rogers}, A.~E.~E., {et~al.} 2008, Nature,
  455, 78

\bibitem[{{Falcke} {et~al.}(1998){Falcke}, {Goss}, {Matsuo}, {Teuben}, {Zhao},
  \& {Zylka}}]{Falcke1998}
{Falcke}, H., {Goss}, W.~M., {Matsuo}, H., {et~al.} 1998, ApJ, 499, 731

\bibitem[{{Falcke} \& {Markoff}(2013)}]{FalckeMarkoff2013}
{Falcke}, H. \& {Markoff}, S.~B. 2013, Classical and Quantum Gravity, 30,
  244003

\bibitem[{{Falcke} {et~al.}(2000){Falcke}, {Melia}, \& {Agol}}]{Falcke2000}
{Falcke}, H., {Melia}, F., \& {Agol}, E. 2000, ApJ L, 528, L13

\bibitem[{{Fish} {et~al.}(2011){Fish}, {Doeleman}, {Beaudoin}, {Blundell},
  {Bolin}, {Bower}, {Chamberlin}, {Freund}, {Friberg}, {Gurwell}, {Honma},
  {Inoue}, {Krichbaum}, {Lamb}, {Marrone}, {Moran}, {Oyama}, {Plambeck},
  {Primiani}, {Rogers}, {Smythe}, {SooHoo}, {Strittmatter}, {Tilanus}, {Titus},
  {Weintroub}, {Wright}, {Woody}, {Young}, \& {Ziurys}}]{Fish2011}
{Fish}, V.~L., {Doeleman}, S.~S., {Beaudoin}, C., {et~al.} 2011, ApJ L, 727,
  L36

\bibitem[{{Fish} {et~al.}(2016){Fish}, {Johnson}, {Doeleman}, {Broderick},
  {Psaltis}, {Lu}, {Akiyama}, {Alef}, {Algaba}, {Asada}, {Beaudoin},
  {Bertarini}, {Blackburn}, {Blundell}, {Bower}, {Brinkerink}, {Cappallo},
  {Chael}, {Chamberlin}, {Chan}, {Crew}, {Dexter}, {Dexter}, {Dzib}, {Falcke},
  {Freund}, {Friberg}, {Greer}, {Gurwell}, {Ho}, {Honma}, {Inoue}, {Johannsen},
  {Kim}, {Krichbaum}, {Lamb}, {Le{\'o}n-Tavares}, {Loeb}, {Loinard},
  {MacMahon}, {Marrone}, {Moran}, {Mo{\'s}cibrodzka}, {Ortiz-Le{\'o}n},
  {Oyama}, {{\"O}zel}, {Plambeck}, {Pradel}, {Primiani}, {Rogers}, {Rosenfeld},
  {Rottmann}, {Roy}, {Ruszczyk}, {Smythe}, {SooHoo}, {Spilker}, {Stone},
  {Strittmatter}, {Tilanus}, {Titus}, {Vertatschitsch}, {Wagner}, {Wardle},
  {Weintroub}, {Woody}, {Wright}, {Yamaguchi}, {Young}, {Young}, {Zensus}, \&
  {Ziurys}}]{Fish2016}
{Fish}, V.~L., {Johnson}, M.~D., {Doeleman}, S.~S., {et~al.} 2016, ApJ, 820,
  90

\bibitem[{{Fraga-Encinas} {et~al.}(2016){Fraga-Encinas}, {Mo{\'s}cibrodzka},
  {Brinkerink}, \& {Falcke}}]{Fraga-Encinas2016}
{Fraga-Encinas}, R., {Mo{\'s}cibrodzka}, M., {Brinkerink}, C., \& {Falcke}, H.
  2016, AAP, 588, A57

\bibitem[{{Ghez} {et~al.}(2008){Ghez}, {Salim}, {Weinberg}, {Lu}, {Do}, {Dunn},
  {Matthews}, {Morris}, {Yelda}, {Becklin}, {Kremenek}, {Milosavljevic}, \&
  {Naiman}}]{Ghez2008}
{Ghez}, A.~M., {Salim}, S., {Weinberg}, N.~N., {et~al.} 2008, ApJ, 689, 1044

\bibitem[{{Gillessen} {et~al.}(2009){Gillessen}, {Eisenhauer}, {Fritz},
  {Bartko}, {Dodds-Eden}, {Pfuhl}, {Ott}, \& {Genzel}}]{Gillessen2009}
{Gillessen}, S., {Eisenhauer}, F., {Fritz}, T.~K., {et~al.} 2009, ApJ L, 707,
  L114

\bibitem[{{Goddi} {et~al.}(2016){Goddi}, {Falcke}, {Kramer}, {Rezzolla},
  {Brinkerink}, {Bronzwaer}, {Deane}, {De Laurentis}, {Desvignes}, {Davelaar},
  {Eisenhauer}, {Eatough}, {Fraga-Encinas}, {Fromm}, {Gillessen}, {Grenzebach},
  {Issaoun}, {Jan{\ss}en}, {Konoplya}, {Krichbaum}, {Laing}, {Liu}, {Lu},
  {Mizuno}, {Moscibrodzka}, {M{\"u}ller}, {Olivares}, {Porth}, {Pfuhl}, {Ros},
  {Roelofs}, {Schuster}, {Tilanus}, {Torne}, {van Bemmel}, {van Langevelde},
  {Wex}, {Younsi}, \& {Zhidenko}}]{bhc_rev}
{Goddi}, C., {Falcke}, H., {Kramer}, M., {et~al.} 2016, ArXiv e-prints
  1606.08879

\bibitem[{{Gravity Collaboration} {et~al.}(2018){Gravity Collaboration},
  {Abuter}, {Amorim}, {Anugu}, {Baub{\"o}ck}, {Benisty}, {Berger}, {Blind},
  {Bonnet}, {Brandner}, {Buron}, {Collin}, {Chapron}, {Cl{\'e}net}, {Coud{\'e}
  Du Foresto}, {de Zeeuw}, {Deen}, {Delplancke-Str{\"o}bele}, {Dembet},
  {Dexter}, {Duvert}, {Eckart}, {Eisenhauer}, {Finger}, {F{\"o}rster
  Schreiber}, {F{\'e}dou}, {Garcia}, {Garcia Lopez}, {Gao}, {Gendron},
  {Genzel}, {Gillessen}, {Gordo}, {Habibi}, {Haubois}, {Haug}, {Hau{\ss}mann},
  {Henning}, {Hippler}, {Horrobin}, {Hubert}, {Hubin}, {Jimenez Rosales},
  {Jochum}, {Jocou}, {Kaufer}, {Kellner}, {Kendrew}, {Kervella}, {Kok},
  {Kulas}, {Lacour}, {Lapeyr{\`e}re}, {Lazareff}, {Le Bouquin}, {L{\'e}na},
  {Lippa}, {Lenzen}, {M{\'e}rand}, {M{\"u}ler}, {Neumann}, {Ott}, {Palanca},
  {Paumard}, {Pasquini}, {Perraut}, {Perrin}, {Pfuhl}, {Plewa}, {Rabien},
  {Ram{\'\i}rez}, {Ramos}, {Rau}, {Rodr{\'\i}guez-Coira}, {Rohloff}, {Rousset},
  {Sanchez-Bermudez}, {Scheithauer}, {Sch{\"o}ller}, {Schuler}, {Spyromilio},
  {Straub}, {Straubmeier}, {Sturm}, {Tacconi}, {Tristram}, {Vincent}, {von
  Fellenberg}, {Wank}, {Waisberg}, {Widmann}, {Wieprecht}, {Wiest},
  {Wiezorrek}, {Woillez}, {Yazici}, {Ziegler}, \& {Zins}}]{Gravity2018}
{Gravity Collaboration}, {Abuter}, R., {Amorim}, A., {et~al.} 2018, AAP, 615,
  L15

\bibitem[{{Gravity collaboration} {et~al.}(2018){Gravity collaboration},
  {Abuter}, {Amorim}, {Baub\"ock}, {Berger}, {Bonnet}, {Brandner}, {Cl\'enet},
  {Coud\'e du Foresto}, {de Zeeuw}, {Deen}, {Dexter}, {Duvert}, {Eckart},
  {Eisenhauer}, {F\"orster Schreiber}, {Garcia}, {Gao}, {Gendron}, {Genzel},
  {Gillessen}, {Guajardo}, {Habibi}, {Haubois}, {Henning}, {Hippler},
  {Horrobin}, {Huber}, {Jim\'enez-Rosales}, {Jocou}, {Kervella}, {Lacour},
  {Lapeyr\`ere}, {Lazareff}, {Le Bouquin}, {L\'ena}, {Lippa}, {Ott}, {Panduro},
  {Paumard}, {Perraut}, {Perrin}, {Pfuhl}, {Plewa}, {Rabien},
  {Rodr\'iguez-Coira}, {Rousset}, {Sternberg}, {Straub}, {Straubmeier},
  {Sturm}, {Tacconi}, {Vincent}, {von Fellenberg}, {Waisberg}, {Widmann},
  {Wieprecht}, {Wiezorrek}, {Woillez}, \& {Yazici}}]{Gravity2018-2}
{Gravity collaboration}, {Abuter}, R., {Amorim}, A., {et~al.} 2018, AAP

\bibitem[{{Greisen}(2003)}]{Greisen2003_AIPS}
{Greisen}, E.~W. 2003, Information Handling in Astronomy - Historical Vistas,
  285, 109

\bibitem[{{Gwinn} {et~al.}(2014){Gwinn}, {Kovalev}, {Johnson}, \&
  {Soglasnov}}]{Gwinn2014_scatteringsubstructure}
{Gwinn}, C.~R., {Kovalev}, Y.~Y., {Johnson}, M.~D., \& {Soglasnov}, V.~A. 2014,
  ApJ L, 794, L14

\bibitem[{{Johnson} {et~al.}(2015){Johnson}, {Fish}, {Doeleman}, {Marrone},
  {Plambeck}, {Wardle}, {Akiyama}, {Asada}, {Beaudoin}, {Blackburn},
  {Blundell}, {Bower}, {Brinkerink}, {Broderick}, {Cappallo}, {Chael}, {Crew},
  {Dexter}, {Dexter}, {Freund}, {Friberg}, {Gold}, {Gurwell}, {Ho}, {Honma},
  {Inoue}, {Kosowsky}, {Krichbaum}, {Lamb}, {Loeb}, {Lu}, {MacMahon},
  {McKinney}, {Moran}, {Narayan}, {Primiani}, {Psaltis}, {Rogers}, {Rosenfeld},
  {SooHoo}, {Tilanus}, {Titus}, {Vertatschitsch}, {Weintroub}, {Wright},
  {Young}, {Zensus}, \& {Ziurys}}]{Johnson2015}
{Johnson}, M.~D., {Fish}, V.~L., {Doeleman}, S.~S., {et~al.} 2015, Science,
  350, 1242

\bibitem[{{Johnson} \& {Gwinn}(2015)}]{JohnsonGwinn2015}
{Johnson}, M.~D. \& {Gwinn}, C.~R. 2015, ApJ, 805, 180

\bibitem[{{Johnson} {et~al.}(2018){Johnson}, {Narayan}, {Psaltis}, {Blackburn},
  {Kovalev}, {Gwinn}, {Zhao}, {Bower}, {Moran}, {Kino}, {Kramer}, {Akiyama},
  {Dexter}, {Broderick}, \& {Sironi}}]{Johnson2018}
{Johnson}, M.~D., {Narayan}, R., {Psaltis}, D., {et~al.} 2018, ApJ, 865, 104

\bibitem[{{Krichbaum} {et~al.}(1998){Krichbaum}, {Graham}, {Witzel}, {Greve},
  {Wink}, {Grewing}, {Colomer}, {de Vicente}, {Gomez-Gonzalez}, {Baudry}, \&
  {Zensus}}]{Krichbaum1998}
{Krichbaum}, T.~P., {Graham}, D.~A., {Witzel}, A., {et~al.} 1998, AAP, 335,
  L106

\bibitem[{{Lo} {et~al.}(1998){Lo}, {Shen}, {Zhao}, \& {Ho}}]{Lo1998}
{Lo}, K.~Y., {Shen}, Z.-Q., {Zhao}, J.-H., \& {Ho}, P.~T.~P. 1998, ApJ L, 508,
  L61

\bibitem[{{Lu} {et~al.}(2011{\natexlab{a}}){Lu}, {Krichbaum}, {Eckart},
  {K{\"o}nig}, {Kunneriath}, {Witzel}, {Witzel}, \& {Zensus}}]{Lu2011_sgra}
{Lu}, R.-S., {Krichbaum}, T.~P., {Eckart}, A., {et~al.} 2011{\natexlab{a}},
  AAP, 525, A76

\bibitem[{{Lu} {et~al.}(2018){Lu}, {Krichbaum}, {Roy}, {Fish}, {Doeleman},
  {Johnson}, {Akiyama}, {Psaltis}, {Alef}, {Asada}, {Beaudoin}, {Bertarini},
  {Blackburn}, {Blundell}, {Bower}, {Brinkerink}, {Broderick}, {Cappallo},
  {Crew}, {Dexter}, {Dexter}, {Falcke}, {Freund}, {Friberg}, {Greer},
  {Gurwell}, {Ho}, {Honma}, {Inoue}, {Kim}, {Lamb}, {Lindqvist}, {Macmahon},
  {Marrone}, {Mart{\'{\i}}-Vidal}, {Menten}, {Moran}, {Nagar}, {Plambeck},
  {Primiani}, {Rogers}, {Ros}, {Rottmann}, {SooHoo}, {Spilker}, {Stone},
  {Strittmatter}, {Tilanus}, {Titus}, {Vertatschitsch}, {Wagner}, {Weintroub},
  {Wright}, {Young}, {Zensus}, \& {Ziurys}}]{Lu2018}
{Lu}, R.-S., {Krichbaum}, T.~P., {Roy}, A.~L., {et~al.} 2018, ApJ, 859, 60

\bibitem[{{Lu} {et~al.}(2011{\natexlab{b}}){Lu}, {Krichbaum}, \&
  {Zensus}}]{Lu2011_nrao530}
{Lu}, R.-S., {Krichbaum}, T.~P., \& {Zensus}, J.~A. 2011{\natexlab{b}}, MNRAS,
  418, 2260

\bibitem[{{Mart{\'{\i}}-Vidal} {et~al.}(2012){Mart{\'{\i}}-Vidal}, {Krichbaum},
  {Marscher}, {Alef}, {Bertarini}, {Bach}, {Schinzel}, {Rottmann}, {Anderson},
  {Zensus}, {Bremer}, {Sanchez}, {Lindqvist}, \& {Mujunen}}]{MartiVidal2012}
{Mart{\'{\i}}-Vidal}, I., {Krichbaum}, T.~P., {Marscher}, A., {et~al.} 2012,
  AAP, 542, A107

\bibitem[{{Mo{\'s}cibrodzka} {et~al.}(2014){Mo{\'s}cibrodzka}, {Falcke},
  {Shiokawa}, \& {Gammie}}]{Moscibrodzka2014}
{Mo{\'s}cibrodzka}, M., {Falcke}, H., {Shiokawa}, H., \& {Gammie}, C.~F. 2014,
  AAP, 570, A7

\bibitem[{{Ortiz-Le{\'o}n} {et~al.}(2016){Ortiz-Le{\'o}n}, {Johnson},
  {Doeleman}, {Blackburn}, {Fish}, {Loinard}, {Reid}, {Castillo}, {Chael},
  {Hern{\'a}ndez-G{\'o}mez}, {Hughes}, {Le{\'o}n-Tavares}, {Lu}, {Monta{\~n}a},
  {Narayanan}, {Rosenfeld}, {S{\'a}nchez}, {Schloerb}, {Shen}, {Shiokawa},
  {SooHoo}, \& {Vertatschitsch}}]{Ortiz2016}
{Ortiz-Le{\'o}n}, G.~N., {Johnson}, M.~D., {Doeleman}, S.~S., {et~al.} 2016,
  ApJ, 824, 40

\bibitem[{{Psaltis} {et~al.}(2015){Psaltis}, {{\"O}zel}, {Chan}, \&
  {Marrone}}]{Psaltis2015}
{Psaltis}, D., {{\"O}zel}, F., {Chan}, C.-K., \& {Marrone}, D.~P. 2015, ApJ,
  814, 115

\bibitem[{{Rauch} {et~al.}(2016){Rauch}, {Ros}, {Krichbaum}, {Eckart},
  {Zensus}, {Shahzamanian}, \& {Mu{\v z}i{\'c}}}]{Rauch2016}
{Rauch}, C., {Ros}, E., {Krichbaum}, T.~P., {et~al.} 2016, AAP, 587, A37

\bibitem[{{Reid}(2009)}]{Reid2009}
{Reid}, M.~J. 2009, International Journal of Modern Physics D, 18, 889

\bibitem[{{Shen}(2006)}]{Shen2006}
{Shen}, Z.-Q. 2006, Journal of Physics Conference Series, 54, 377

\bibitem[{{Shen} {et~al.}(2005){Shen}, {Lo}, {Liang}, {Ho}, \&
  {Zhao}}]{Shen2005}
{Shen}, Z.-Q., {Lo}, K.~Y., {Liang}, M.-C., {Ho}, P.~T.~P., \& {Zhao}, J.-H.
  2005, Nature, 438, 62

\bibitem[{{Shepherd}(1997)}]{Shepherd1997}
{Shepherd}, M.~C. 1997, in Astronomical Data Analysis Software and Systems VI,
  ed. {G.~Hunt \& H.~Payne}, ASP Conf.\ Proc. 125, 77

\end{thebibliography}

\begin{appendix}

\section{Modelfitting technique}\label{app:fitting}

In the modelfitting algorithm, we select at random 2 independent closure amplitudes out of 6 possible ones for each quadrangle and integration time to be used in the model fitting procedure. We perform the model fitting of the independent closure amplitudes by using a gradient descent method, where the source model parameters are iteratively altered to give successively better (lower) $\chi^2$-scores until convergence is reached. The 2D Gaussian model we employ has 3 free parameters: major axis size (FWHM), minor axis size (FWHM) and the position angle on the sky of the major axis. For every bootstrapping realization, a random point in the 3D model parameter space is initially chosen as a starting point, from a flat distribution using upper limits for the major and minor axes sizes of 400 $\mu$as (and lower limits of 0 $\mu$as) to ensure rapid convergence. Initial coarse step sizes are 50 $\mu$as for both major and minor axes, and 0.1 radians for the position angle. For the parameter starting point, as well as for its neighbours along all dimensions (each one step size removed from the initial point along one parameter axis), the $\chi^2$ scores are calculated and the lowest-scoring point in the resulting set is taken as the starting point for the next iteration. This sequence of steps is repeatedly performed until the best-fitting model parameters coincide with the starting point for that iteration (indicating a local optimum has been reached at that parameter resolution), after which the step sizes for all parameters are reduced and the algorithm continues until the minimum step sizes for all parameters are reached. To verify that the general nature of the $\chi^2$ landscape is conducive to this iterative method, and to ensure that the algorithm would not get stuck in a local optimum rather than the global optimum, we have mapped out the $\chi^2$ scores over the full 3D parameter space at a low resolution for the original full set of closure amplitudes. This investigation suggested that the $\chi^2$-score varies smoothly over the full parameter space, revealing the presence of a single global optimum.\\

\end{appendix}

\end{document}